\documentclass{emulateapj}

\usepackage{amsmath,amssymb}
\usepackage{natbib}
\usepackage{epsfig,graphics}
\usepackage{epstopdf} 
\usepackage{ulem}

\newcommand{\lsim}{\raise0.3ex\hbox{$<$}\kern-0.75em{\lower0.65ex\hbox{$\sim$}}}
\newcommand{\gsim}{\raise0.3ex\hbox{$>$}\kern-0.75em{\lower0.65ex\hbox{$\sim$}}}

\begin{document}
\slugcomment{Draft Version; not for circulation}
\title{Exoplanet Transit Spectroscopy Using WFC3:  WASP-12 b, WASP-17 b, and WASP-19 b}

\author{Avi M. Mandell\altaffilmark{1,7}, Korey Haynes\altaffilmark{1,2}, Evan Sinukoff\altaffilmark{3}, Nikku Madhusudhan\altaffilmark{4}, Adam Burrows\altaffilmark{5}, Drake Deming\altaffilmark{6} 
}
\altaffiltext{1}{Solar System Exploration Division, NASA Goddard Space Flight Center, Greenbelt, MD 20771, USA}
\altaffiltext{2}{School of Physics, Astronomy, and Computational Sciences, George Mason University, Fairfax, VA 22030, USA}
\altaffiltext{3}{Institute for Astronomy, University of Hawaii, Honolulu, HI 96822, USA}
\altaffiltext{4}{Yale Center for Astronomy and Astrophysics, Yale University, New Haven, CT 06511, USA}
\altaffiltext{5}{Department of Astrophysical Sciences, Princeton University, Princeton, NJ 08544, USA}
\altaffiltext{6}{Department of Astronomy, University of Maryland, College Park, MD 20742, USA}
\altaffiltext{7}{Corresponding Email:  Avi.Mandell@nasa.gov}

\begin{abstract}
We report analysis of transit spectroscopy of the extrasolar planets WASP-12 b, WASP-17 b, and WASP-19 b using the Wide Field Camera 3 on the HST. We analyze the data for a single transit for each planet using a strategy similar in certain aspects to the techniques used by \citet{Berta:2012ff}, but we extend their methodology to allow us to correct for channel- or wavelength-dependent instrumental effects by utilizing the band-integrated time series and measurements of the drift of the spectrum on the detector over time.  We achieve almost photon-limited results for individual spectral bins, but the uncertainties in the transit depth for the the band-integrated data are exacerbated by the uneven sampling of the light curve imposed by the orbital phasing of HST's observations.  Our final transit spectra for all three objects are consistent with the presence of a broad absorption feature at 1.4\,$\mu$m potentially due to water. However, the amplitude of the absorption is less than that expected based on previous observations with {\it Spitzer}, possibly due to hazes absorbing in the NIR or non-solar compositions. The degeneracy of models with different compositions and temperature structures combined with the low amplitude of any features in the data preclude our ability to place unambiguous constraints on the atmospheric composition without additional observations with WFC3 to improve the S/N and/or a comprehensive multi-wavelength analysis.
\end{abstract}



\section{Introduction}
\label{I}

Over the past decade there has been significant progress in characterizing exoplanets orbiting a wide variety of nearby stars, including the first detections of light emitted by an exoplanet \citep{Charbonneau:2005be,Deming:2005fg}, the first spectrum of an exoplanet \citep{Richardson:2007eu, Grillmair:2007ee, Swain:2008fm}, the first phase curve for an exoplanet \citep{Knutson:2007bl}, the first detection of haze in an exoplanetary atmosphere \citep{Pont:2008ft}, and tentative constraints claimed for the water, methane, carbon monoxide and carbon dioxide abundances in several exoplanetary atmospheres \citep{Grillmair:2008je, Swain:2008fm, Swain:2009ds, Swain:2009ed, Madhusudhan:2009gd, Madhusudhan:2011kw}. Almost 100 transiting exoplanets with $V_{star} < 12$ have been discovered to date, many with multi-band photometry from both space and ground-based observatories.  We are firmly in the era of exoplanet characterization, and yet the sparse data available for each planet has resulted in more questions than answers.

The Wide Field Camera 3 (WFC3) on the {\it Hubble Space Telescope} (HST) provides the potential for spectroscopic characterization of molecular features in exoplanet atmospheres, a capability that has not existed in space since the demise of NICMOS on HST and the IRS on {\it Spitzer}.  WFC3 is an optical/NIR camera capable of slitless grism spectroscopy, with wavelength coverage in the the IR spanning between 0.8 and 1.7\,$\mu$m.  Studies of exoplanets have focused on using the G141 grism, the long-wavelength dispersion element on the infrared channel that covers the wavelength range 1.1\,$\mu$m to 1.7\,$\mu$m at a maximum resolving power of 130 at 1.4\,$\mu$m \citep{Dressel2012handbook}.  This region spans both the major bands of water between 1.3 and 1.5\,$\mu$m as well as another water band at 1.15\,$\mu$m, and bands of a few other molecular species.  Observations measuring flux within NIR water bands are impossible from the ground due to the extinction and variability caused by water vapor in Earth's atmosphere; WFC3 therefore represents the only current platform for measuring absorption and/or emission from water in exoplanet atmospheres.

In this paper we present WFC3 observations of three transiting ``hot Jupiter" exoplanets --- WASP-12 b, WASP-17 b, and WASP-19 b --- during transit of the host star.  Two of these data sets, for WASP-17 b and WASP-19 b, were observed as part of a large HST program to examine single transits and occultations from 16 hot Jupiters (P.I. D. Deming), while the data for the transit of WASP-12 b were taken as part of a single-object campaign (P.I. M. Swain) and first analyzed in \citet{Swain:2013hn}.  All three planets orbit extremely close to their parent star and have large atmospheric scale heights, making them excellent targets for transmission spectroscopy.   WASP-12 b and WASP-17 b (as well as WASP-19 b to a lesser extent) belong to a class of ``bloated" or ``inflated" planets, which have significantly larger radii than would be predicted from traditional evolutionary models \citep{Burrows:2000ho, Guillot:2002cq}.  WASP-17 b is also in a retrograde orbit compared to the rotation of its host star \citep{Anderson:2010dx, Bayliss:2010ch, Triaud:2010hr}, while WASP-12 b and WASP-19 b appear to be in prograde orbits \citep{Albrecht:2012hk,Hellier:2011hf}.  Retrograde orbits have commonly been interpreted as evidence that the planet was forced into a highly inclined and eccentric orbit through planet-planet scattering \citep{Rasio:1996ie,Weidenschilling:1996ig} or the Kozai mechanism \citep{Fabrycky:2007jh}, and was subsequently re-circularized through dissipation of orbital energy by tides \citep{Jackson:2008ip}. The extremely short orbit of WASP-19 b also argues for tidal decay after scattering \citep{Hellier:2011hf}. In the tidal decay scenario the large radii of the planets could be due to internal dissipation of tidal energy during orbital circularization \citep{Bodenheimer:2001eu}. However, based on recent models by \citet{Ibgui:2009fw}, \citet{Anderson:2011hf} conclude that any transient tidal heating produced during circularization of the orbit of WASP-17 b would have dissipated by the time the planet reached its current orbit, making the planet's large radius unsustainable.  Other theories for the misalignment of the stellar rotation and the planet's orbit do not require a previous eccentric orbit and tidal re-circularization \citep{Rogers:2012dm}, and a number of other theories for the heating mechanisms required to produce large planetary radii have been proposed, including ``kinetic heating" due to the dissipation of wind energy deep in the atmosphere \citep{Guillot:2002cq} and Ohmic dissipation \citep{Batygin:2010dz}; therefore the dynamical origin of these extremely hot and inflated giant planets is still highly uncertain.

In principle, understanding the atmospheric composition of hot Jupiters can help constrain their formation and dynamical histories.  Unfortunately, observational studies have produced conflicting results regarding the atmospheric compositions of several hot Jupiters, including WASP-12 b and WASP-19 b.  \citet{Madhusudhan:2011kw} first raised the possibility of a non-solar abundance in the atmosphere of WASP-12 b using occultation measurements in four {\it Spitzer} photometric bands \citep{Campo:2011fx} and three ground-based NIR photometric bands \citep{Croll:2011jt} to constrain the carbon-to-oxygen ratio to super-solar values, possibly greater than unity.  Similar {\it Spitzer} and ground-based measurements for WASP-19 b were consistent with both solar and super-solar C/O models \citep{Anderson:2013dx}, raising the possibility of a population of carbon-rich hot Jupiters.  However, \citet{Crossfield:2012gm} recently re-analyzed the {\it Spitzer} data for WASP-12 b in light of the discovery of a faint candidate companion imaged by \citet{Bergfors:2013de}, concluding that the dilution-corrected {\it Spitzer} and ground-based photometry can be fit by solar-metallicity models with almost isothermal temperature structures.

While transmission spectroscopy only weakly constrains the overall temperature structure of a transiting exoplanet, it can place strong constraints on the presence of molecular features in absorption through the limb of the planet, thereby constraining the atmospheric composition.  Models by \citet{Madhusudhan:2012ga} suggest that spectral features of H$_2$O and hydrocarbons (e.g. CH$_4$, HCN, and C$_2$H$_2$) will change drastically with different C/O values, and the WFC3 bandpass covers several of these features.  In this paper we present our data reduction and analysis of the three transits, including our analysis of contamination from nearby sources and our strategy to compensate for the significant instrumental systematics in much of the WFC3 data, and conclude with preliminary constraints on the atmospheric composition and structure of the three planets.


\section{Observations}
\label{O}

The observations of WASP-17 and WASP-19 analyzed here were conducted between June and July of 2011, while the observations of WASP-12 were obtained in April of 2011. Observation dates and exposure information are listed in Table~\ref{obs}.  The observations were taken with the G141 grism on WFC3's infrared channel, providing slitless spectra covering the wavelength range 1.1\,$\mu$m to 1.7\,$\mu$m at a maximum resolving power of 130 at 1.4\,$\mu$m \citep{Dressel2012handbook}.  Dithering was avoided to minimize variations in pixel-to-pixel sensitivity. The ``spatial scanning" mode suggested as a strategy to increase efficiency and decrease persistence for bright objects \citep{2012McCullough_spatscan} was not used since it had not been developed at the time of observation.  Each target was allocated 4--5 HST orbits, each lasting 90 minutes followed by 45 minute gaps due to Earth occultations of the telescope.  This was sufficient to cover a single transit while including some out-of-transit data as well.

\begin{deluxetable}{cccc}
\tablecaption{Observation Parameters}
\tabletypesize{\scriptsize}
\tablewidth{0pt}
\tablehead{\colhead{} &
                    \colhead{WASP-12} &
                    \colhead{WASP-17} &
                    \colhead{WASP-19}}
\startdata
Date of Observation  & 2011-04-12   &  2011-07-08  & 2011-07-01\\
Integration Time         & 7.624   &  12.795  & 21.657\\
Subarray Mode          & 256   &  512  & 128\\
CALWF3 version       & 2.7 & 2.3  & 2.3\\
NSamp                        & 3 & 16  & 5\\
Timing Sequence      & SPARS10 & RAPID  & SPARS10\\
Peak Pixel Value\tablenotemark{1} & 38,000 & 64,000 & 73,000 \\
\enddata
\tablenotetext{1}{The number of electrons recorded at the peak of the spectral distribution in a single exposure.}
\label{obs}
\end{deluxetable}

The IR channel of the WFC3 instrument uses a 1024 x 1024 pixel detector array, but smaller sub-arrays can be downloaded to decrease the readout time and increase the exposure cadence.  Additionally, there are two possible sampling sequences:  RAPID sampling, which reads as quickly as possible (limited only by the readout time per sub-array) in order to maximize sampling for short exposures of bright targets, and SPARS sampling, which takes two quick reads and then spaces reads linearly, to allow ``sampling up the ramp", or SUTR.  RAPID sampling naturally has shorter readout times for each sub-array size but imposes a maximum integration time, while the SPARS10 sampling sequence has a minimum exposure time of $\sim$7 sec but no maximum. 

Observations of WASP-17 were taken using the 512 x 512 sub-array with 16 non-destructive reads per exposure and sampled using the RAPID sampling sequence.  This resulted in a total integration time of 12.795 seconds per exposure and 27 exposures per orbit, with a total of 131 exposures taken over five HST orbits.  Observations of WASP-19 were taken using the 128 x 128 sub-array mode with 5 non-destructive reads per exposure, sampled with the SPARS10 sequence. This resulted in an integration time of 21.657 seconds and 70 exposures per orbit, with a total of 274 exposures taken over four orbits. The WASP-12 data utilized the 256 x 256 sub-array mode with 3 non-destructive reads per exposure, leading to an integration time of 7.624 seconds and 99 exposures per orbit, with 484 exposures taken over five HST orbits. We discuss the implications of each sub-array size with respect to systematic trends in \S\ref{Sys}.

\section{Data Reduction}
\label{DR}

\subsection{Image Files:  {\tt .flt} vs {\tt .ima}}

The WFC3 {\tt calwf3} calibration pipeline processes the raw detector output into two calibrated files per exposure:  a file comprising the individual, non-destructive reads (called the {\tt .ima} file) and a single final image produced by determining the flux rate by fitting a line to the individual read-out values for each pixel (called the {\tt .flt} file).  The calibration steps implemented for the {\tt .ima} files include reference pixel subtraction, zero-read and dark current subtraction, and a non-linearity correction; additional corrections applied using SUTR fitting for the {\tt .flt} files include cosmic-ray and bad-pixel flagging and gain calibration.  While it would seem that the {\tt .flt} files would be the best choice for analysis, an analysis of the noise characteristics for each data type revealed that time series extracted from the {\tt .flt} files have an rms that is on average 1.3$\times$ greater than time series created from the {\tt .ima} files.  It is unclear where this difference originates, though it is probably due to inaccurate cosmic ray flagging for very bright sources (STScI WFC3 Support, private communication); we therefore decided to determine our own flux values for each pixel directly from the {\tt .ima} files and essentially re-create our own {\tt .flt} files as a starting point for our analysis (this method was also advocated by \citet{Swain:2013hn} for similar reasons).

\begin{figure}[htb]
\centering
{
\includegraphics[width=85mm]{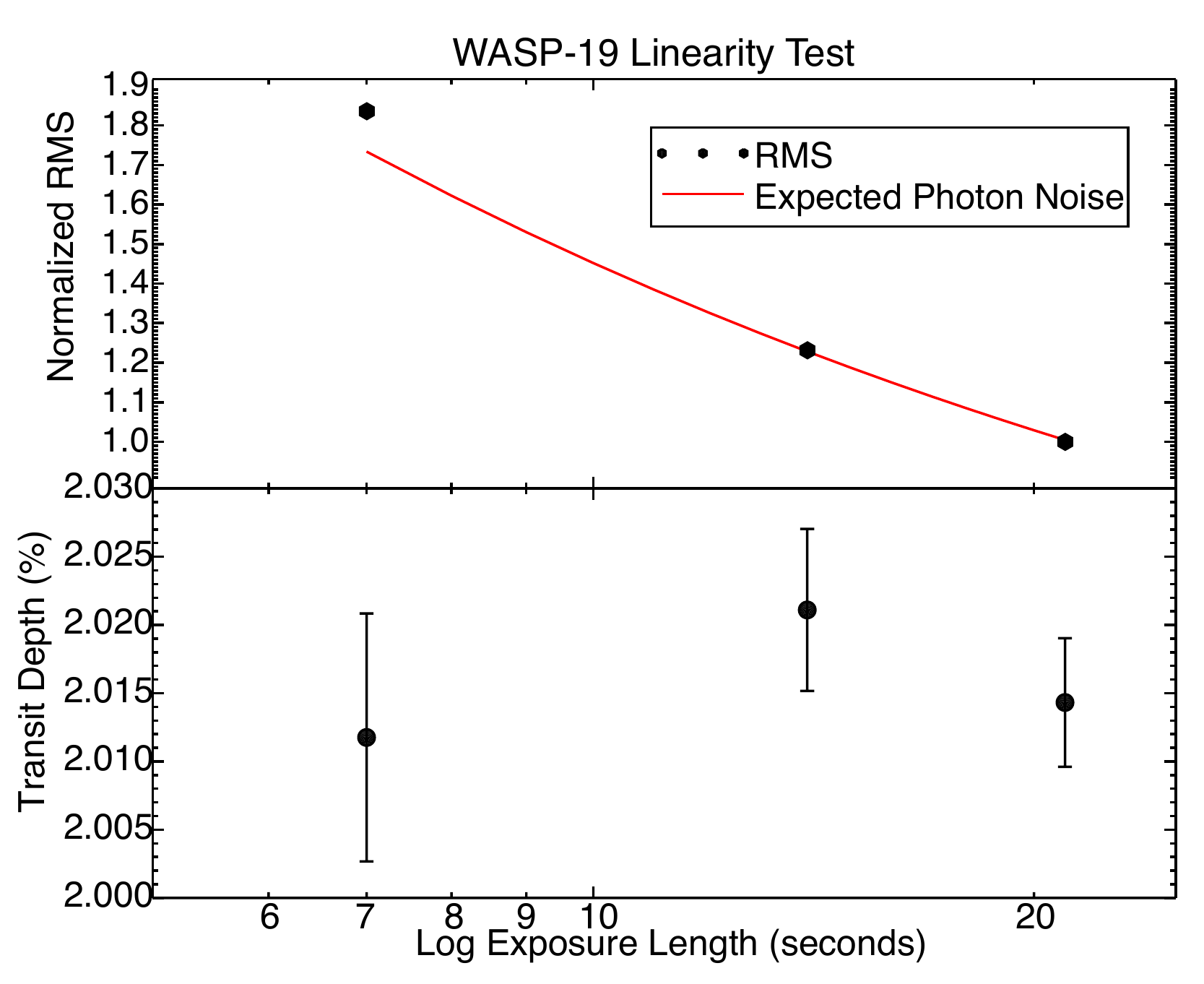}
 \caption{Top: The normalized rms compared to expected photon noise for a band-integrated light curve for WASP-19 created using different individual reads. Bottom: Fitted transit depth for each read. The rms follows the photon-limited trend except for the first point, which most likely reflects read noise; the best-fit values are the same within uncertainties. }
 \label{fig:linearity}
}
\end{figure}

Though the {\tt .ima} files include a linearity correction, the exposures for some our objects approached or exceeded the established linearity limit for WFC3 and we therefore examined our data for signs of any remaining non-linearity.  The WFC3 detector generally remains linear up to 78K e$^{-}$ (WFC3 Handbook); however, \citet{Swain:2013hn} suggest that known WFC3 issues with systematic increases in counts between buffer downloads (see \S\ref{Sys}) may be present when count levels exceed 40K DN, or the equivalent of 100K e$^{-}$.  Our peak counts reach a maximum of 73K e$^{-}$for WASP-19, with lower values for our other targets (see Table~\ref{obs}); we therefore chose WASP-19 to examine linearity.  WASP-19 only has a total of 4 SUTR measurements; in Figure \ref{fig:linearity} we show that the normalized rms of our band-integrated light curve follows the expected decrease for a photon-limited case.  We also examined the linearity of each channel separately, in order to search for correlations with the final transit depth.  Deviations from linearity were $\sim$0.8\% on average, but the channel-to-channel differences were only $\sim$0.1\% and would affect the transit depths for individual channels by only $\sim$20 ppm, far below our uncertainty limits. After binning up channels, this effect would be even less; we therefore did not use any additional linearity correction.

\subsection{Spectral Extraction}
\label{SE}
The unique requirements of time-series photometry of bright sources necessitated the development of a custom-designed data reduction process for WFC3 exoplanet data. A data reduction package called aXe \citep{Kummel:2009dn} exists for analyzing WFC3 data, but this software was designed with dithered observations in mind, and we used the package only for generating a wavelength solution and nominal extraction box sizes since the package incorporates the most recent configuration files for the instrument. An object list was first generated by SExtractor \citep{Bertin:1996ww}, which uses the direct image to find the position of each source. aXe then calculates the trace and wavelength solution for each source, and produces FITS files with an extracted box from each grism image (with the extension {\tt .stp}) and a 1D spectrum (with the extension {\tt .spc}) from which we extract the wavelength solution. For simplicity, we assumed that each pixel in a column has the same wavelength solution; measurements of the center of a Gaussian fit to the dispersion in the y direction showed that it changes by less than 0.02 pixel along the length of the spectra for all of our objects, so this assumption is valid.  We also checked our wavelength solutions against the standard WFC3 sensitivity function to confirm accuracy for all sources.

We retrieved the coordinates for the extraction box from the headers of the {\tt .stp} files, but we decided to expand the number of rows included in the extraction box from 15 pixels to 20 pixels to ensure that we included as much of the wings of the spatial PSF as possible while avoiding any possible contamination from background sources. We also trimmed the extraction boxes to exclude regions of the spectrum with low S/N, keeping the central 112 pixels of each spectrum.

\subsection{Flat Field, Background Subtraction, Bad Pixel and Cosmic Ray Correction}
\label{BPC}

The {\bf calwf3} pipeline does not correct for pixel-to-pixel variations in grism images, but a flat-field cube is provided on the WFC3 website \citep{Kuntschner:2008tq}. Each extension of the cube contains a coefficient, developed by ground tests, that can be fed into a polynomial function as follows, where $x = (\lambda-\lambda_{min})/(\lambda_{max}-\lambda_{min})$ and $\lambda$ is the wavelength of pixel (i,j):

\begin{equation} f(i,j,x)=a_{0} + a{1}*x+a_{2}*x^{2}+...a{i}*x^{n} \end{equation}

\begin{figure}[htb]
\centering
{
\includegraphics[width=85mm]{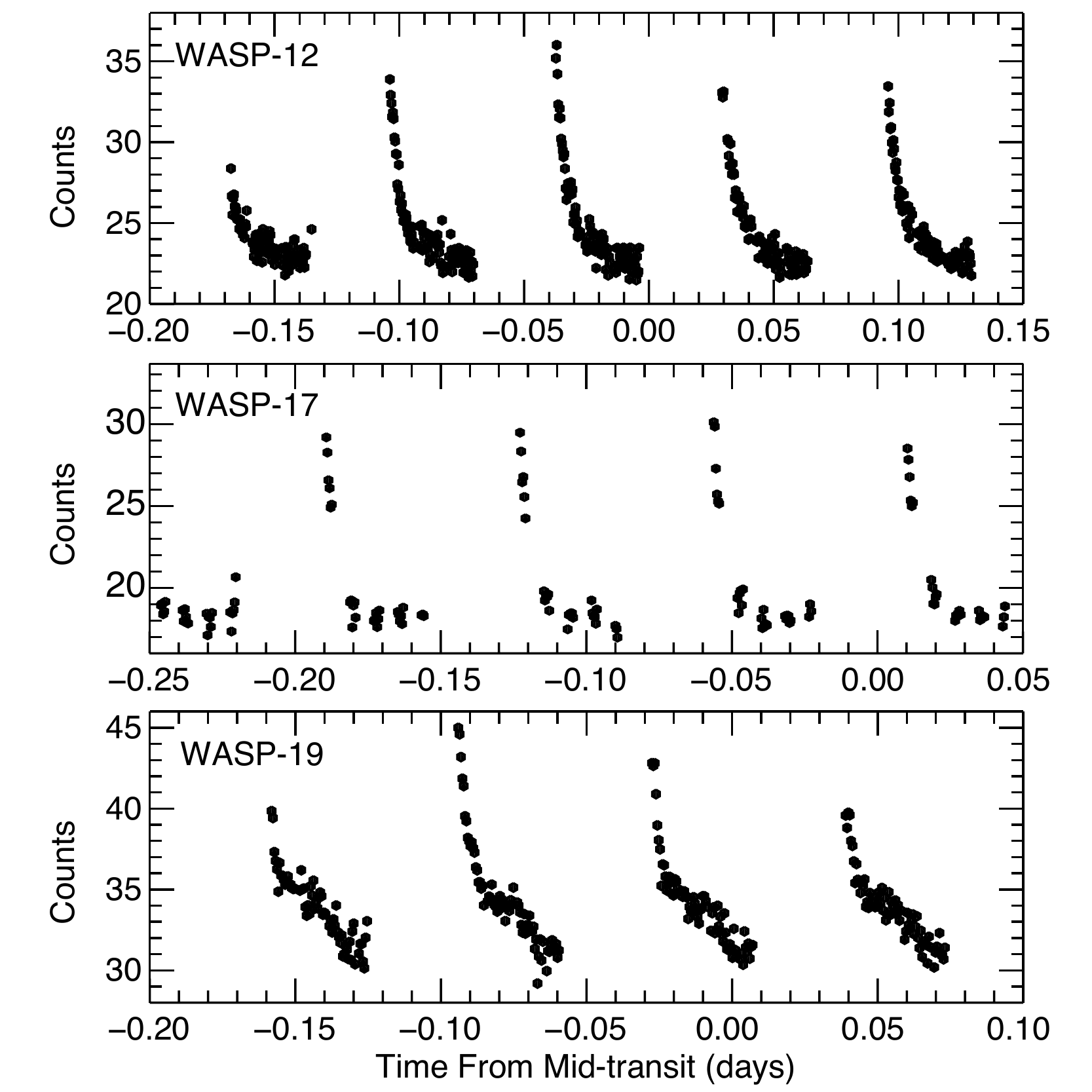}
 \caption{Background levels in counts for each target, given as a function of time from mid-transit. The drop in flux on a per-orbit basis is similar for each target, indicating that instrumental effects such as thermal variations during orbit are the likely cause.}
 \label{fig:background}
}
\end{figure}

This polynomial gives the value of the flat field at each pixel in the extraction region, and we divided this flat field from our data.  We also subtracted an average background flux from each spectral channel by using nearby uncontaminated regions of each image. These background regions cover the same wavelength space (extent in the x direction) as our science box, and are placed as far from the primary source as possible, leaving only a few pixels to guard against edge effects. We then averaged these background rows in the y direction, and subtracted this background spectra from each row of our science box. The average value of the background region was $\sim$15 - 35 $e^{-}$, but for each source background counts drop quickly at the beginning of each orbit and then continue to decrease slowly over the orbital duration (see Figure \ref{fig:background}).  The pattern is very similar in each channel, and is most likely due to thermal variations during the orbit.

To identify pixels that are either permanently bad or contaminated by cosmic rays, we employed several different bad pixel identification strategies. First, to find individual pixels in individual images that were contaminated by cosmic rays or sensitivity variations, we created a 3D image cube and examined each pixel in the 2D images over time; any single-image pixels that were  $> 6\sigma$ higher than the median of their counterparts in time were flagged.  We found 62 bad pixels for WASP-12, 30 for WASP-17, and 120 for WASP-19. We corrected most of these pixels through spatial interpolation in their individual frames; however, the linear interpolation that we used to correct bad pixels would clearly not be effective within the region covered by the stellar PSF due to the rapid change in flux  across pixels in the spatial direction.  Bad pixels within the PSF would also clearly have severe effects on the time series even if they were corrected, and we therefore left these pixels uncorrected.

We then summed over the spatial dimension of the corrected cube yielding a 2D (wavelength, time) array and normalized this array in both the spectral and temporal dimensions, allowing us to remove the band-integrated transit signal and the stellar and instrumental spectral characteristics.  This allowed us to identify both bad spectral channels in individual images as well as individual images and/or channels that showed increased noise or unusual characteristics.  Through this analysis we found 20 individual bad data points for WASP-12, 6 for WASP-17, and 16 for WASP-19, which we corrected by linear interpolation in the spectral dimension.  Additionally, we identified several spectral channels in each data set whose time series showed a significantly higher rms scatter compared with the rest of the channels; we removed 2 channels for WASP-12, 4 channels for WASP-17, and 1 channel for WASP-19 from further analysis as well.

\section{Analysis and Results}
\label{An}

\subsection{Instrumental Systematics}
\label{Sys}
\begin{figure}[htb]
\centering
{
\includegraphics[width=85mm]{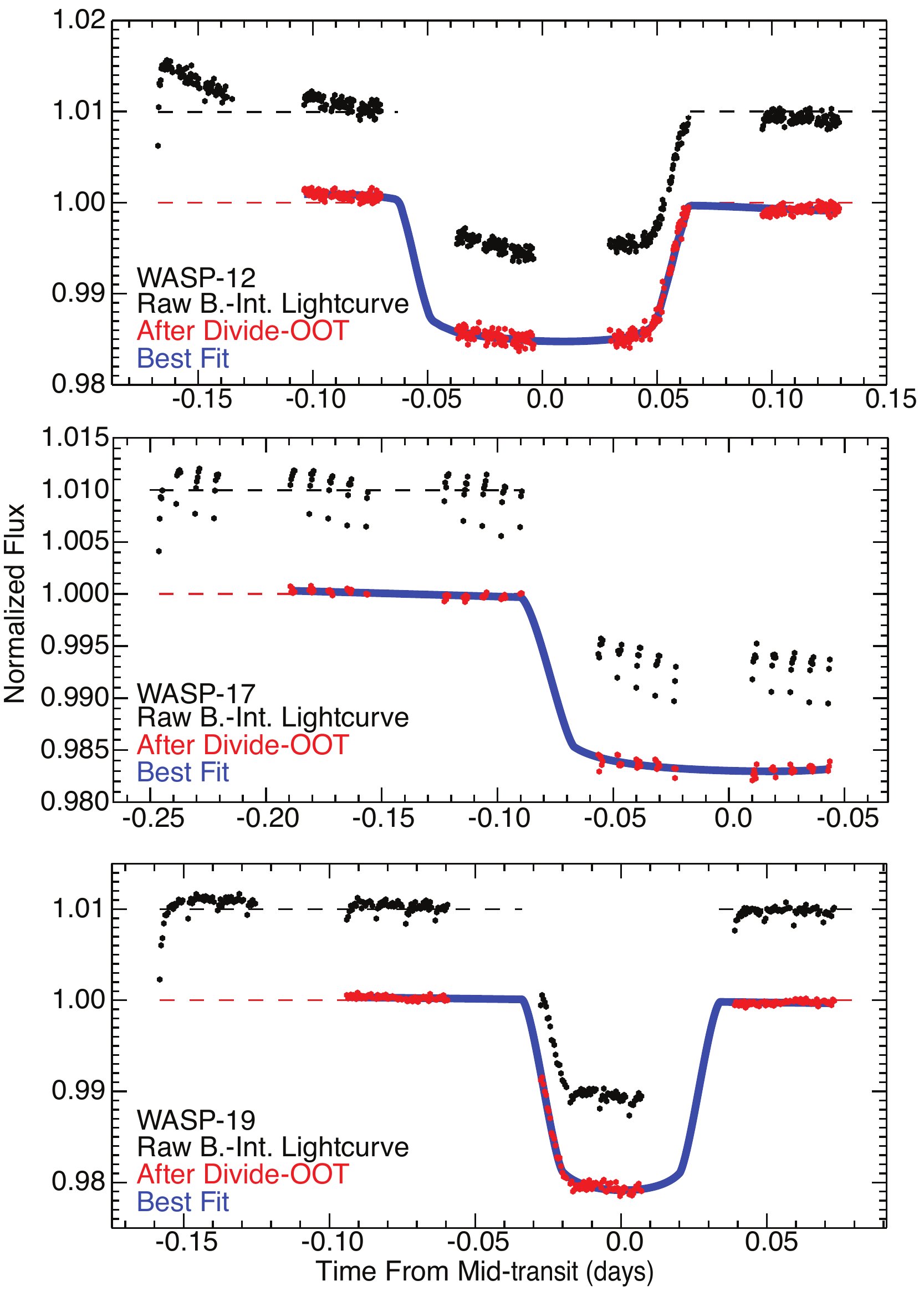}
 \caption{The combined-light time series for each source, before and after removing systematic trends.  The presence of an intra-orbit pattern is easily identified for WASP-17 and WASP-19, repeating after every buffer read-out, but less obvious for WASP-12.  After excluding the first orbit which is inconsistent with the others due to telescope settling, we removed the trends using the divide-oot method devised by \citet{Berta:2012ff}.}
 \label{fig:rawlight}
}
\end{figure}

Two out of our three data sets show strong systematic trends with time, which can be attributed to various instrumental effects, and have been seen in previous observations \citep{Berta:2012ff,Swain:2013hn}. The most obvious trend is the pattern of increasing counts after each data buffer download to the solid-state drive, possibly due to the use of charge-flush mode during the download \citep{Swain:2013hn}. Depending on how quickly the count level stabilizes, this pattern can resemble a ``ramp" (continually increasing until the next buffer download) or a ``hook" (increasing for several exposures and then stabilizing). The effect may be associated with the well-known persistence effects inherent in HgCdTe detectors in general \citep{Smith:2008by} and confirmed in WFC3 in particular \citep{McCullough:2008tl}, but the relationship to the data buffer downloads suggests a connection to the data storage devices.  \citet{Swain:2013hn} performed an exhaustive analysis of the buffer-ramp effects in a number of different sources, and suggest that a smaller sub-array size, a fewer number of non-destructive reads, and a lower illumination level will decrease or eliminate the effect; for reference, we list the relevant attributes for each target in Table~\ref{obs}. The band-integrated light curves (Figure~\ref{fig:rawlight}) for the three objects we analyze here follow this general relationship - WASP-12 (intermediate array size, 3 reads, low peak pixel flux) has no buffer-ramp effect, while WASP-17 (large array size, 16 reads, high peak pixel flux) has a very steep ramp-up with no apparent stabilization before the next buffer dump. WASP-19 (small array size, 5 reads, high peak pixel flux) displays a shape intermediate between the two (a ``hook"-like shape). We do not attempt a more detailed analysis of the cause of the buffer-ramp effects; we find that the divide-oot method developed by \citet{Berta:2012ff} is sufficient to remove the effect almost completely in the band-integrated light curve provided sufficient out-of-transit data is available. We also see a visit-long decrease in flux; this effect has been noted in previous WFC3 analyses and may be due to a slow dissipation of persistence charge, and we correct for it using a linear trend component in our transit model fit. As noted in previous work, the first orbit for each target showed substantially higher scatter than all other orbits, and we do not use this orbit in our band-integrated divide-oot analysis; however, for our wavelength-dependent analysis we use a relative-depth analysis (see \S\ref{WLCF} and \S\ref{BLCF} for a detailed description), and with this fitting strategy we are able to incorporate the first noisier orbit.

\begin{figure}[htb]
\centering
{
\includegraphics[width=85mm]{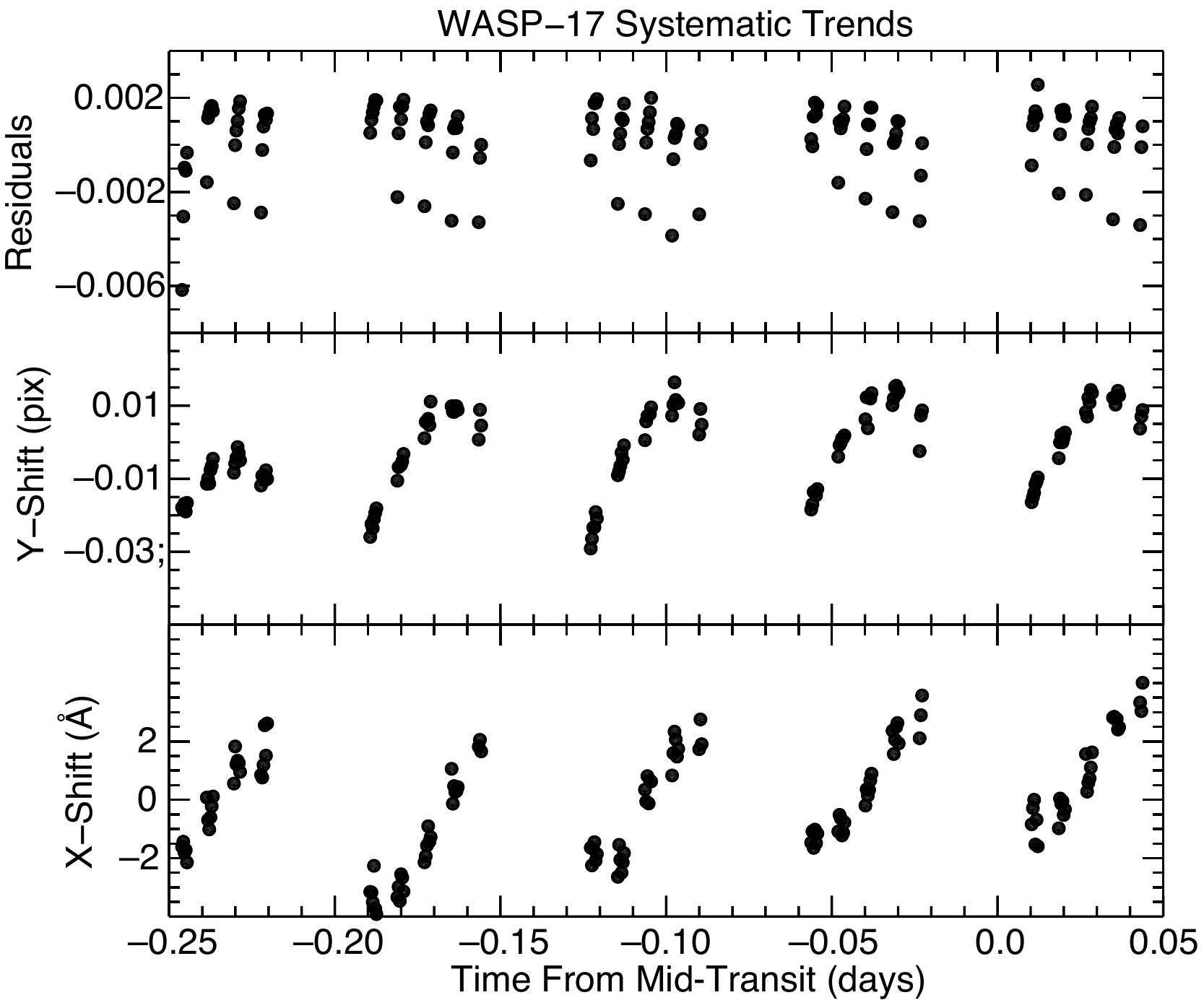}
 \caption{Top: The residuals of the combined-light fit for WASP-17, after subtracting our best-fit model.  Middle: The shift in the position of the spatial profile of WASP-17 over time, in pixels.  The vertical shift was calculated by fitting a Gaussian to sum of the spectral box in the spectral direction.  Bottom: the shift in the position of the spectral profile over time, in pixels.  The horizontal shift was calculated by measuring the change in flux over the edges of the spectrum and deriving the required shift of the spectral sensitivity function (see \S\ref{Sys}).}
 \label{fig:xyshifts}
}
\end{figure}

For each image, we also calculated the shift in the vertical (i.e. spatial) and horizontal (i.e. spectral) directions referenced to the first exposure in the time series.  This allowed us to correct for any modulation in channel flux due to undersampling of the spatial PSF and/or spectral features. Since the FWHM of the PSF is $\sim$3 pixels, any vertical shifts can have a significant effect on the illumination of individual rows, and a similar effect can occur due to features in the stellar spectrum or the WFC3 sensitivity function that are several pixels wide. However, the shifts we measure are only a fraction of a pixel (see Figure~\ref{fig:xyshifts}) and the motion of a pixel across the spatial PSF or a spectral feature will be extremely small, creating a change in flux that is essentially linear.  We can therefore decorrelate this effect against a scaled measurement of the image motion in each direction.  

We measured the vertical shift by first summing our extraction box in the wavelength direction to get a 1D array of the flux absorbed by each row of the detector for each exposure and then fitting a Gaussian to those arrays to determine the change in the location of the peak of the flux distribution from the first exposure. A precise measurement of the horizontal shift (i.e. the spectral shift) across all exposures was more difficult to calculate, since the sensitivity function of the grism does not allow for an analytical fit.  We first attempted to cross-correlate the spectra against each other, but the scatter in the resulting measurements was too high to be useful. We then decided to utilize the edges of the spectrum where the sensitivity function of the detector rises and falls rapidly, and a small change in pixel position will have a strong effect on the illumination of each pixel. We fit a line to the slope for the same pixels at the edge of the spectrum for each exposure, and used the intercept of this fit to determine the shift of each spectrum in relation to the first exposure; the values from the fit to both the short-wavelength and long-wavelength edge of each spectrum were averaged to decrease the effective uncertainty of the measurement.  In Figure~\ref{fig:xyshifts} the vertical and horizontal shifts, as well as the final band-integrated residuals after subtracting a light curve model, are plotted for WASP-17 as an example. All of the variables change relatively coherently within an orbit, and then reset at the beginning of the next orbit. 

\subsection{Background Source Correction}
\label{FFBC}

We also examined each object for contamination from background sources.  Due to the slitless design of WFC3, spectra from background sources can be shifted both spatially and spectrally compared with the science target.  In particular, a nearby background source or companion was discovered for WASP-12 \citep{Bergfors:2013de} and more recently confirmed to be a double star \citep{Bechter:2013uv}; the close companions have been shown to significantly affect the mid-IR photometry of this source with {\it Spitzer} \citep{Crossfield:2012gm}.  After averaging all of the images for each source, we examined each combined image by eye for evidence of background contamination, and then used a vertical profile cut to further constrain the amplitude and location of any identified sources.  For WASP-19 there were no additional sources, and for WASP-17 the single background source identified nearby was very dim and significantly shifted in the spatial direction from the science target and therefore exterior to our extraction box.  

\begin{figure}[htb]
\centering
{
\includegraphics[width=85mm]{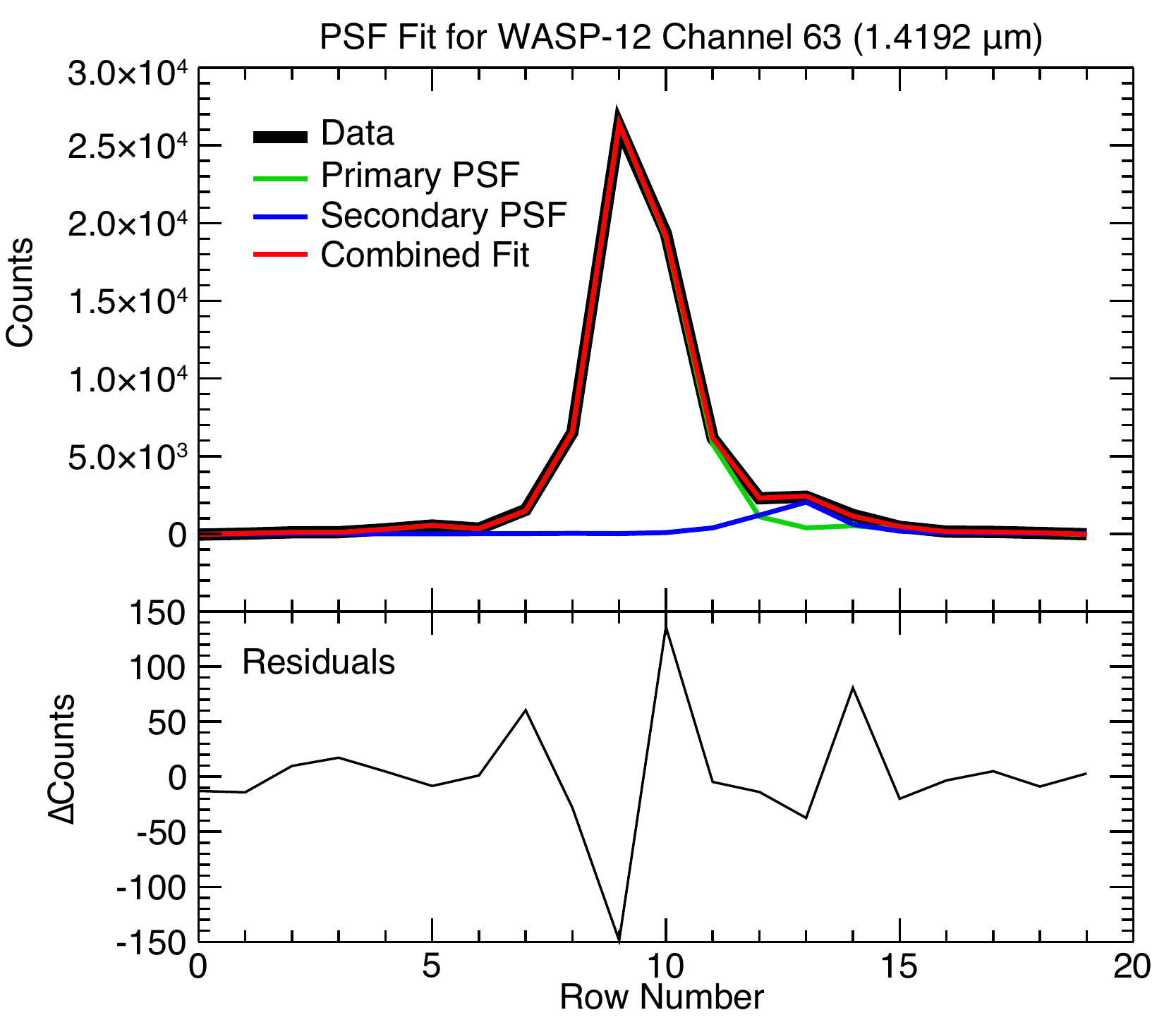}
 \caption{Top: Data and the best-fit PSF model for a single channel for WASP-12, using the template-PSF method. The data are shown in black, the fit to the main peak is shown in green, and the fit to the contamination peak is shown in blue; the combined fit is shown in red. Bottom: Remaining residuals after removing the model; the remaining flux under the region of contamination was used as the uncertainty in the contamination flux.}
 \label{fig:PSF1}
}
\end{figure}

For WASP-12 we identified a relatively bright contamination source very close to the science target; the peak of the spectral profile of the secondary source is located only $\sim$4 pixels away from the peak of the primary stellar PSF in the spatial direction (see Figure~\ref{fig:PSF1}).  This object is most likely the source identified by \citet{Bergfors:2013de} (referred to as Bergfors-6 by \citet{Crossfield:2012gm} and WASP-12 BC by \citet{Bechter:2013uv}); after correcting for a shift of the the secondary source in the spectral direction, the separation between the two sources matches up well with the previous measurements.  As stated above, \citet{Bechter:2013uv} resolved the source into two stars, but in the direct image from HST they are unresolved - the difference in the FWHM of the primary PSF compared to the secondary PSF is only 0.25 pixels. We therefore refer to the combined contamination from the two stars in our data as WASP-12 BC.  \citet{Swain:2013hn} also identified this contamination, and fit the profile of the PSF in the spatial direction by using the PSF shape from separate observations of a reference star; this method has the benefit of providing an empirical PSF shape that can be used for both the brighter primary star as well as the secondary star.  This strategy is slightly complicated in this instance because of the multiplicity of the secondary source, but as stated above, the change in the width of the PSF is extremely small.  The more difficult problem is that the angle of the spectrum on the detector is slightly offset from the horizontal pixel pitch; therefore the PSF changes shape with wavelength, and the primary and secondary point spread functions are sampled differently.

\begin{figure}[htb]
\centering
{
\includegraphics[width=85mm]{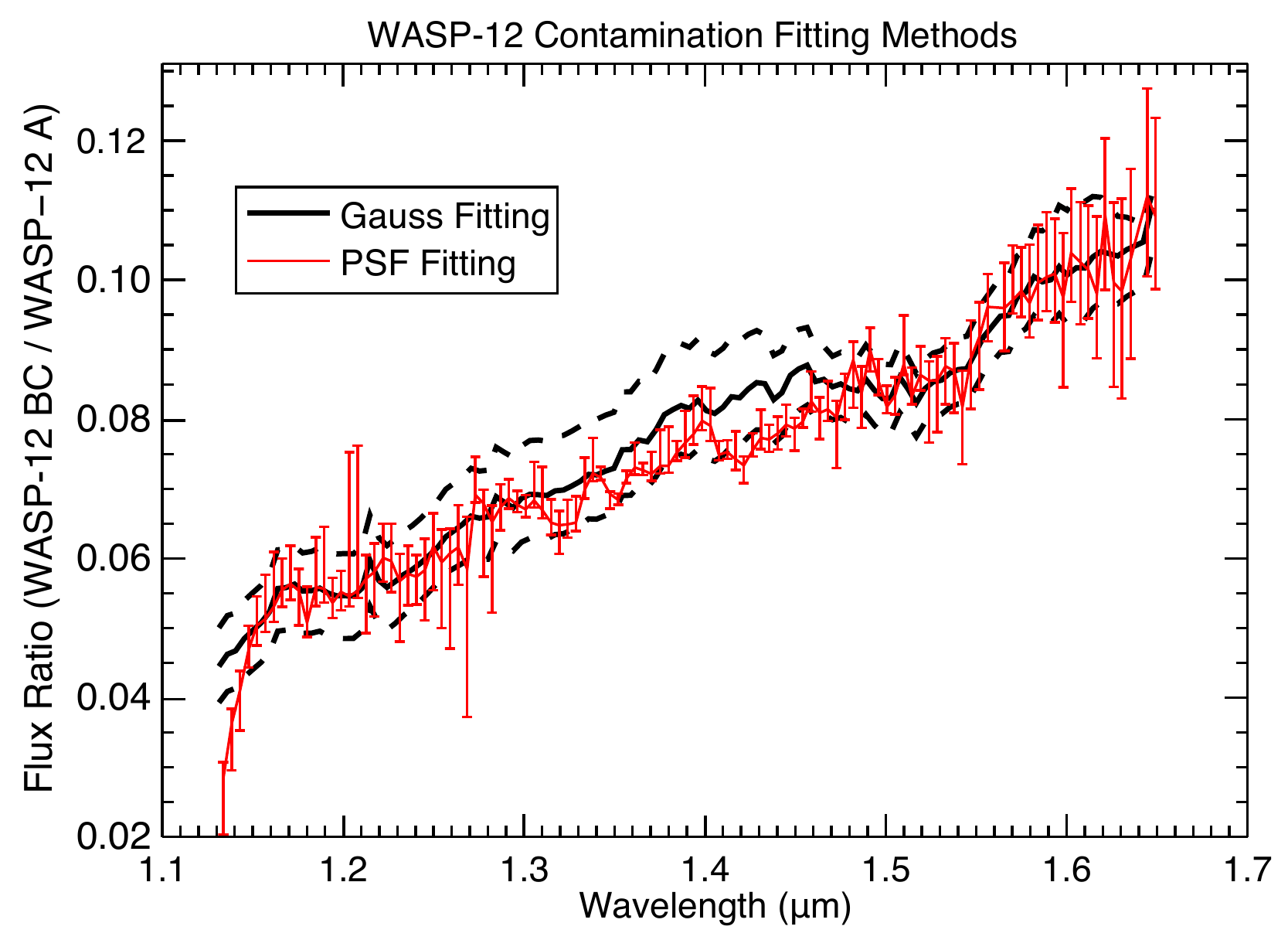}
 \caption{Flux ratio for WASP-12 BC compared to WASP-12 A for our two fitting methods. Gaussian fitting (black) subtracts one Gaussian centered on WASP-12's position, then fits another Gaussian to the residuals, centered on the contaminating source. Template PSF fitting (red) jointly scales two PSFs, using pre-determined columns from WASP-19 as a template. The uncertainties using the template PSF method are much smaller, even with the distortions of the secondary PSF due to the multiplicity of WASP-12 BC.}
 \label{fig:PSF2}
}
\end{figure}

Fortunately our WASP-19 spectrum was also slightly angled on the detector, and since the flux levels remaining in the linear regime we were able to scale individual channels from our WASP-19 data as PSF ``templates" for the WASP-12 channels (as suggested by \citet{Swain:2013hn}). The PSF of WASP-12 BC could also be fit in the same way, albeit with a different initial off-set for the starting template channel. We empirically determined the best-fit template channel off-set for both PSFs, and then performed a least-squares fit for the PSF amplitude of both stars at once. In Figure~\ref{fig:PSF1} we show an example of a fit to one of our WASP-12 channels; the remaining residuals in the region with the contaminating source will be impacted slightly by the distorted PSF of the double stars, so we summed them up to give uncertainties on the fit in the positive and negative directions. In addition to this PSF template strategy, we tested a straightforward sequential Gaussian fitting method, first fitting and subtracting the largest-amplitude signal (from the science target) and then fitting the additional contamination source. However, due to the under-sampling of the spatial PSFs and their overlap between the two sources, there was substantial uncertainty in the fundamental baseline of the individual PSF functions for each source, and considerable residual flux was left over after removing the contribution from both PSFs. In Figure~\ref{fig:PSF2} we plot our spectrum for WASP-12 BC derived from both methods. The results agree extremely well at short and long wavelengths except for an overall offset and some slight discrepancy between 1.35 and 1.45\;$\mu$m; however, the uncertainties are at least a factor of 3 smaller using the template-PSF method, even with the contributing error from the multiplicity of WASP-12 BC.  We therefore adopted the results from the template-PSF fitting method, and corrected the data by subtracting the derived spectrum of the contaminating source from the 1D spectrum at each time step.

\begin{figure}[htb]
\centering
{
\includegraphics[width=90mm]{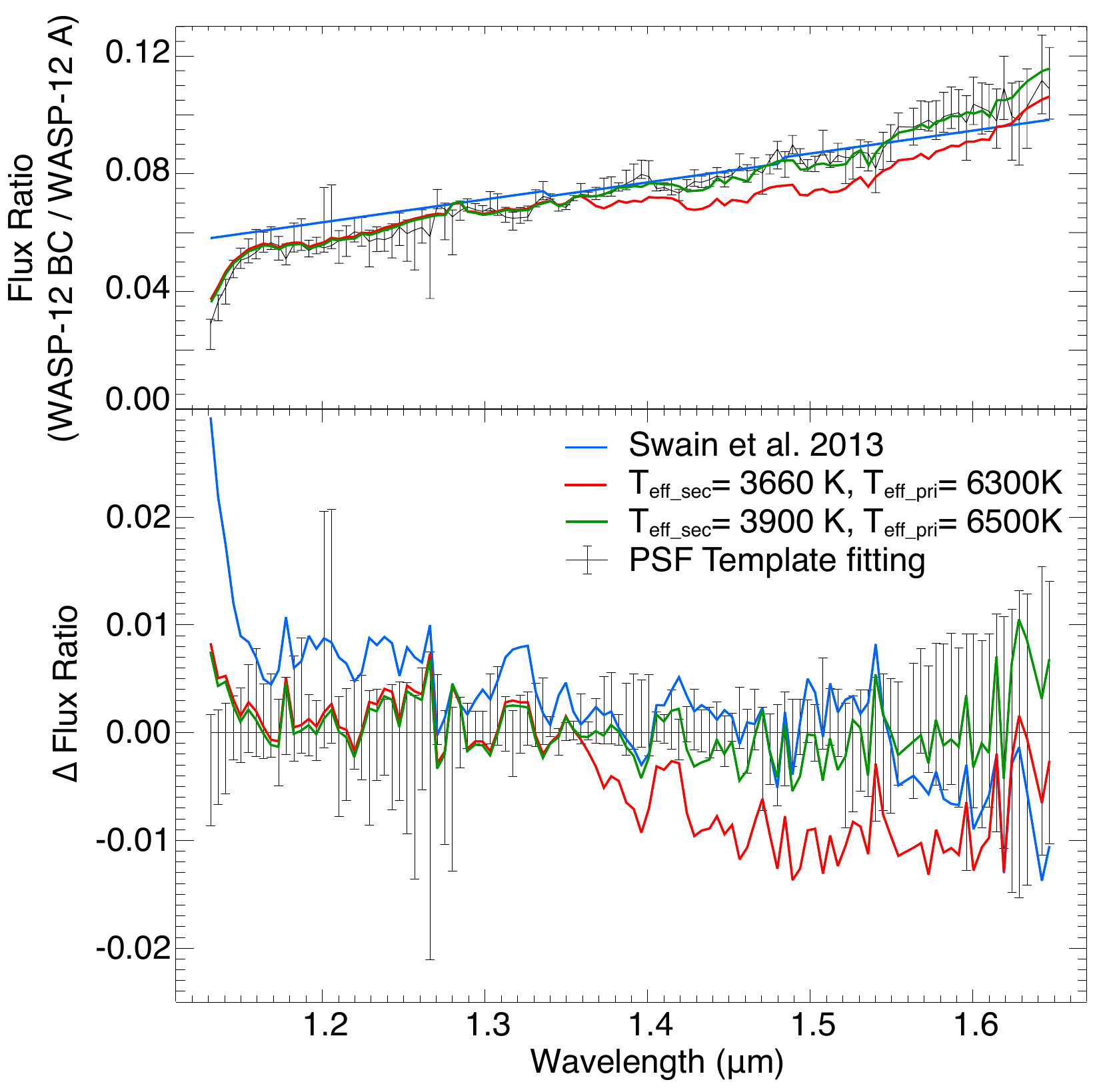}
 \caption{Top: Flux ratio for the contaminating source (WASP-12 BC) from the template PSF fitting method (black), compared with analytical models for the flux ratio bracketing the range of values for the temperatures of WASP-12 and the contaminating source (red and green); an approximation to the same values from \citet{Swain:2013hn} are also plotted (blue). Bottom: the same analyses as above, but both the analytical models and the \citet{Swain:2013hn} results have the values from our fitting subtracted, in order to better show the discrepancies.  The results from our PSF fitting match very closely with the high-temperature limit for the temperatures of both the primary source (WASP-12 A) and the contaminating source (WASP-12 BC); the low-temperature model shows a much larger signature of absorption from water vapor between 1.35 and 1.6\;$\mu$m.  The \citet{Swain:2013hn} results are similar at most wavelengths, but there is a very large discrepancy at the shortest wavelengths.}
 \label{fig:contam_compare}
}
\end{figure}

For comparison, we calculated the expected ratio of the contaminating source to the primary star using stellar atmosphere models from \citet{Castelli:2004ti}, assuming that the contaminating source is the combined light from WASP-12 B and WASP-12 C. \citet{Crossfield:2012gm} determined a spectral type of M0V and an effective temperature between 3600 K and 3900 K for what they believed was a single star, depending on whether purely spectroscopic or a combination of spectroscopic and photometric data were used; for WASP-12, \citet{Hebb:2009fw} determined an effective temperature of 6300$^{200}_{100}$ K. Since \citet{Bechter:2013uv} find that both companions have a similar spectral type and brightness, we can effectively treat them as one source.  We assumed the same metallicity for all the stars, and used the direct image to derive a shift of 331 \AA\ in the spectral direction for the contaminating source. We then scaled the ratio of two stellar models to match our results at 1.6\;$\mu$m. In Figure~\ref{fig:contam_compare} we plot our results from our PSF template method, with two analytic models spanning the range of effective temperatures for WASP-12 A and WASP-12 BC. A lower-temperature model for the combined flux from WASP-12 BC shows a significantly deeper water absorption feature from 1.4 to 1.6\;$\mu$m compared with higher-temperature models, while a higher temperature for WASP-12 A makes a very small change in the overall slope. Our empirical fit to the data agrees very well with a model using a temperature of $\sim$3900 K for WASP-12 BC, which matches well with the M0 spectral type derived by \citet{Bergfors:2013de} and \citet{Crossfield:2012gm} but is inconsistent with the spectral type of M3V determined by \citet{Bechter:2013uv} for both WASP-12 B and C.  

A similar calculation of the contaminating flux was used by \citet{Stevenson:2013wf}; however, they assumed the lower effective temperature for WASP-12 BC from the spectroscopic analysis by \citet{Crossfield:2012gm} {\it a priori}, without attempting to determine the contaminating flux empirically. Alternately, \citet{Swain:2013hn} performed a similar fit to ours, but their results appear to lack the sharp downturn shortwards of 1.15\;$\mu$m and the upturn longwards of 1.55\;$\mu$m that are evident in our results; the slope of their results is also slightly shallower (a linear approximation to their results is plotted in Figure~\ref{fig:contam_compare}; they did not publish their fitted values, but they are close to a single linear trend with a slight decrease between 1.34 and 1.48\;$\mu$m). Given the close similarity between the high-temperature analytical model and our empirical fit to the data, we remain confident that our results are robust. However, it is clear that the choice of the spectral dependence for the dilution by WASP-12 BC has a significant impact on the final results for the spectrum of WASP-12 b; uncertainties of 1\% for the dilution factor for WASP-12 BC will result in a difference of 150 ppm in the final transit depth, which is similar in magnitude to the uncertainties for the transit depths of our individual bins.  We discuss this impact further in \S\ref{disc}.

\subsection{Band-Integrated Transit Curve Fitting}
\label{WLCF}

Our analysis strategy relies on the assumption that almost all of the time-dependent trends present in the band-integrated time series are consistent across wavelength (even if the amplitudes of these trends change), since the systematics are related to either the general exposure parameters (array size, number of read-outs, etc), and/or correlated with the illumination of each pixel. We therefore decided to determine the band-integrated transit curve parameters first, and then use the residuals from this band-integrated fit as a component in our transit model when fitting individual spectral channels (with the amplitude of this component allowed to vary). This method allows us to incorporate any common-mode systematic trends into our fit, providing a more robust measurement of the relative change in transit depth across spectral channels, which is the most important factor when measuring the depth of spectral absorption features. We are also able to include the first orbit for each target into the wavelength-dependent analysis since the higher scatter in this orbit (which has caused most observers to discard it) is common across wavelength and can be removed accurately. We describe the fitting strategy in more detail in \S\ref{BLCF}.

To achieve the best possible fit to the band-integrated light curve prior to fitting individual spectral bins, we utilized the divide-oot method developed by \citet{Berta:2012ff}, which uses the systematics in the out-of-transit data to correct the in-transit data by simply dividing all orbits by an average of the out-of-transit orbits. This method works very well to remove the repeated intra-orbit slope and buffer-ramp effects, which represent the largest instrumental effect in our data. We then fit the corrected light curve with a Markov Chain Monte Carlo (MCMC) routine with a Metropolis-Hastings algorithm within the Gibbs sampler \citep{Ford:2005gi}, using the light curve model from \citet{Mandel:2002bb}, with an additional linear slope term to account for the gradual decrease in flux seen in all WFC3 exoplanet transit data to date.

All of the orbital parameters in our transit light curve model were locked to the literature values (see Table \ref{litvals}), since we are only analyzing single transits and lack full-transit coverage. The only exceptions are the mid-transit time and the two parameters for a quadratic limb darkening law, which we allow to vary under Gaussian priors since we are only analyzing a single transit with incomplete coverage of ingress and egress. For mid-transit times, we calculate the predicted mid-transit time from recent transit observations of our targets in the literature, and propagate the uncertainty on period in time to use as the width of our prior. For limb darkening, we use values calculated by \citet{Claret:2011gy} from analysis of ATLAS models. After selecting for the appropriate stellar parameters, \citet{Claret:2011gy} provide values at the centers of the J and H bands, with a choice between a least-square and flux conservation method. We interpolated between the J and H band points to find the central wavelength of our spectra, and took the average between the two methods as our starting limb-darkening parameter value.  We used the standard deviation between the two methods, multiplied by two, as the width of our priors.

\begin{deluxetable*}{cccc}
\tablecaption{Stellar and Orbital Parameters Used For Model Fitting and Comparison}
\tabletypesize{\scriptsize}
\tablewidth{0pt}
\tablehead{\colhead{Parameters} &
                     \colhead{WASP-12 b\tablenotemark{a}} &
                    \colhead{WASP-17 b\tablenotemark{b}} &
                    \colhead{WASP-19 b\tablenotemark{c}}}
\startdata
Period (days) & 1.09  & 3.73 & 0.789  \\
$i$ ($^{\circ}$) & 82.5 $\pm$ 0.8 & 86.7 $\pm$ 0.500 & 79.5 $\pm$ 0.500 \\
$R_{p}/R^{*}$ & 0.117 $\pm$ 0.00068 & 0.123 $\pm$ 0.037 & 0.139 $\pm$ 0.0457 \\
$T_{c}$  & 55663.199  & 55750.285 & 55743.532 \\
$\mu_1$\tablenotemark{d} & 0.127 $\pm$ 0.0487 & 0.0901 $\pm$ 0.0487 & 0.153 $\pm$ 0.0487 \\
$\mu_2$  & 0.271 $\pm$ 0.0620 & 0.273 $\pm$ 0.0620 & 0.293 $\pm$ 0.0620 \\
$a/R^{*}$ & 3.03 $\pm$ 0.0220 & 6.96 $\pm$ 0.0220 & 3.57 $\pm$ 0.0460 \\
e & 0.0447 $\pm$ 0.00430 & 0.00  & 0.00770 $\pm$ 0.00680 \\
$\omega$ ($^{\circ}$) & 94.4 $\pm$ 0.0300 & 0.00 & 43.0 $\pm$ 67.0 \\

Semi-major axis (AU) & 0.02309 $\pm$ 0.00096 & 0.05105 $\pm$ 0.00128 & 0.01616 $\pm$ 0.00024 \\	
$M_{*}$ (M$_{\odot}$)	 & 1.38 $\pm$ 0.18	&	1.286 $\pm$ 0.079	& 0.904 $\pm$ 0.040	\\	
$M_{p} \times$sin $i$ (M$_{J}$)  & 1.378 $\pm$ 0.181 &	0.477 $\pm$ 0.033	& 1.114 $\pm$ 0.04 \\
Spectral type	 &	G0	&		F4	&	G8V \\	
H-band Magnitude        & 10.228    & 10.319   & 10.602 \\
\lbrack Fe/H\rbrack	 &	0.3 $\pm$ 0.1	&	-0.25 $\pm$ 0.09		& 0.02 $\pm$ 0.09	\\	
\enddata
\label{litvals}
\tablenotetext{a}{Values from \citet{Southworth:2012fv}.}
\tablenotetext{b}{Values from \citet{Maciejewski:2013ip}. }
\tablenotetext{c}{Values from \citet{Lendl:2013jda}. }
\tablenotetext{d}{Values for limb darkening derived from \citet{Claret:2011gy} quadratic limb darkening tables. }
\end{deluxetable*}

\begin{figure}[htb]
\centering
{
\includegraphics[width=85mm]{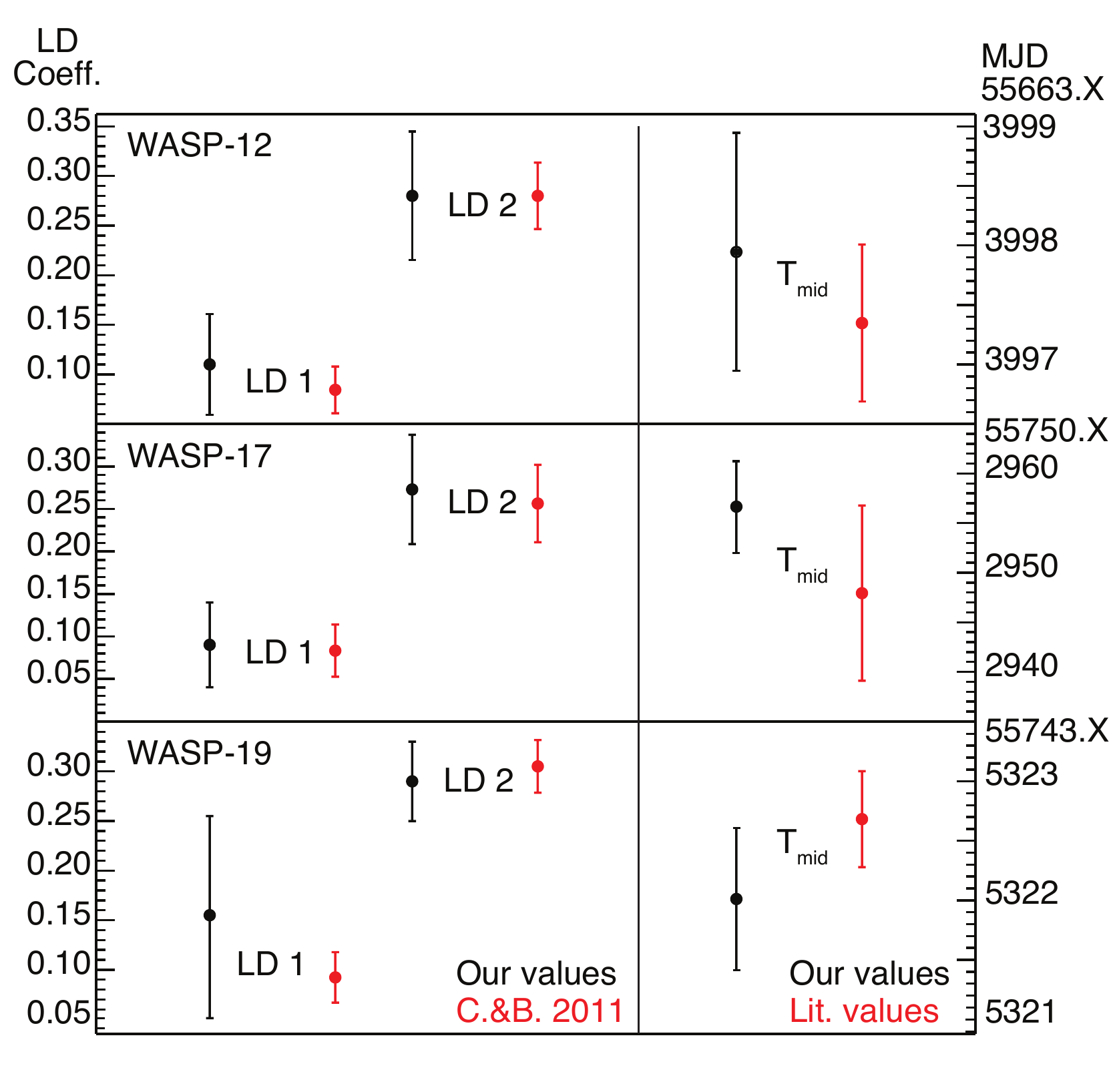}
 \caption{Limb darkening parameters for a quadratic limb darkening law shown as calculated using models from \citet{Claret:2011gy}, and as found by our MCMC routine, using the \citet{Claret:2011gy} models and uncertainties as priors. Our final values match the expected values within uncertainties for all targets.}
 \label{fig:ld}
}
\end{figure}

For each light curve we ran three MCMC chains with 100,000 links for analysis, with an additional initial burn period of 25,000 links.  Our band-integrated time series for each of our targets are shown in Figure~\ref{fig:rawlight}, with the best-fit transit curve overlaid; we tabulate our best-fit orbital parameters in Table \ref{bfvals}. Our best-fit limb darkening parameters compare well with the expected values from \citet{Claret:2011gy}, and best-fit mid-transit times are within the uncertainties based on prior measurements (see Figure~\ref{fig:ld}).

\begin{deluxetable*}{cccc}
\tablecaption{Fitted Parameters From Band-Integrated Time Series}
\tabletypesize{\scriptsize}
\tablewidth{0pt}
\tablehead{\colhead{Parameters} &
                    \colhead{WASP-12 b} &
                     \colhead{WASP-17 b} &
                    \colhead{WASP-19 b}}
\startdata
$R_{p}/R^{*}$ & 0.11895 $\pm$ 0.0013 & 0.12316 $\pm$ 0.00058 & 0.14140 $\pm$ 0.00093 \\
$\mu_1$ & 0.085 $\pm$ 0.024 & 0.083 $\pm$ 0.031 & 0.092 $\pm$ 0.025 \\
$\mu_2$ & 0.281 $\pm$ 0.034 & 0.256 $\pm$ 0.046 & 0.305 $\pm$ 0.027 \\
Mid-Transit (MJD) & 55663.199736 $\pm$ 0.000065 & 55750.294793 $\pm$ 0.00088 & 55743.532268 $\pm$ 0.000040 \\
Slope\tablenotemark{a} & -0.00793 $\pm$ 0.00034 & -0.00578 $\pm$ 0.0010 & -0.00407 $\pm$ 0.00039 \\
\enddata
\label{bfvals}
\tablenotetext{a}{Linear slope has units of normalized flux per day.}
\end{deluxetable*}

\subsubsection{Fitting for a Possible Thermal Contribution and Starspots}
\label{W19anom}

\begin{figure}[htb]
\centering
{
\includegraphics[width=85mm]{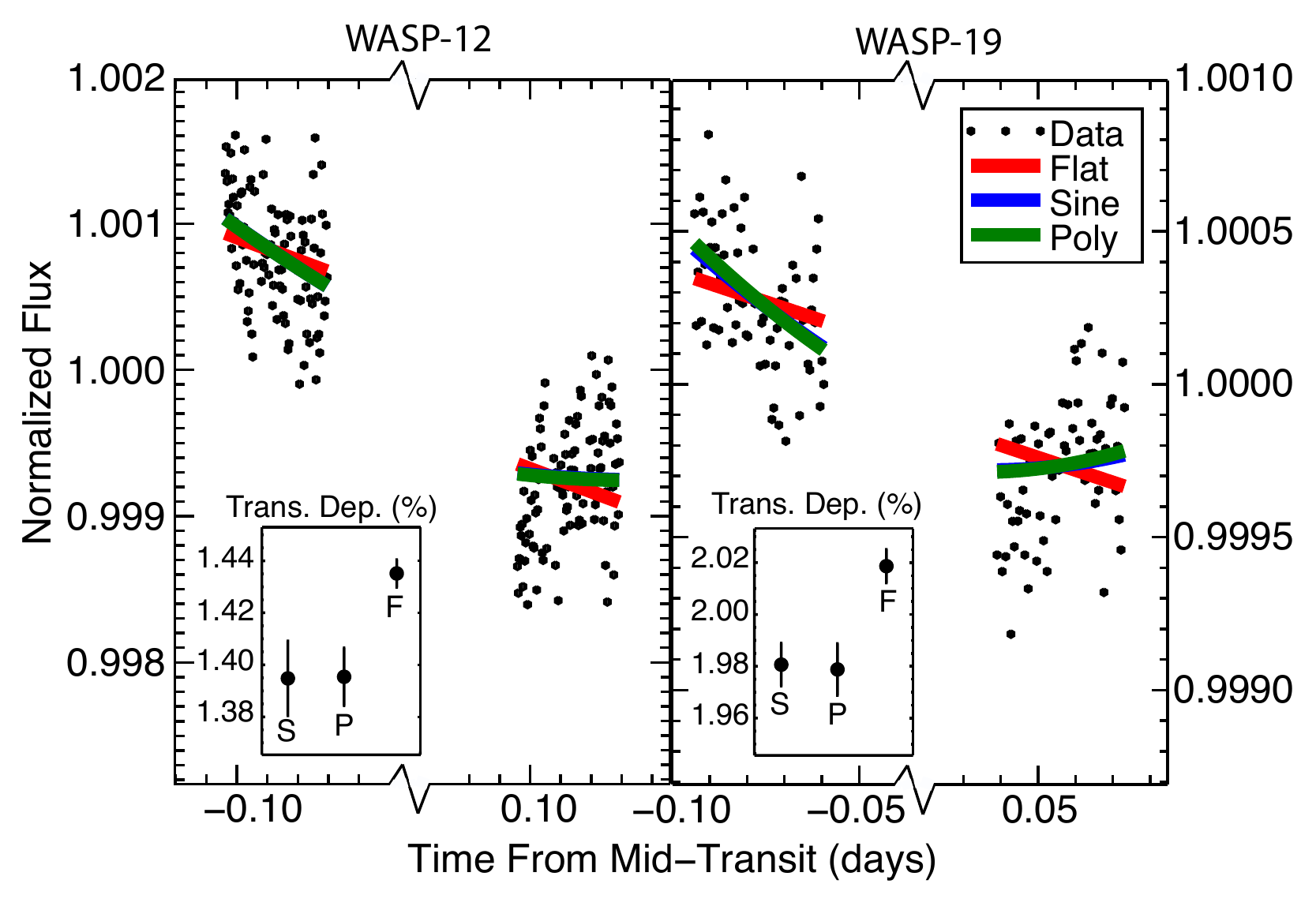}
 \caption{The out-of-transit portions of the band-integrated light curves for WASP-12 (left) and WASP-19 (right), with models including only a linear trend (red) and an additional sinusoidal component (blue) or 2nd-order polynomial function (green) over-plotted.  The best-fit transit depths for each model are also plotted (inset). The addition of sinusoidal or polynomial components produce a marginally better fit, but the improvements are not sufficient to yield a lower BIC.}
 \label{fig:sine}
}
\end{figure}

After fitting the integrated-light time series using the standard transit model, we determined that there appeared to be systematic deviations in the residuals of the out-of-transit orbits for both WASP-12 and WASP-19 as well as the in-transit orbit for WASP-19. The out-of-transit orbits appear to have trends in flux that are not perfectly fit by a single linear slope, with the first orbit having a steeper slope while the last orbit has a shallower slope (see Figure~\ref{fig:sine}). It is difficult to determine the source of these trends due to the limited sampling in orbital phase and the necessity of using the divide-oot correction method, which combines the data from all out-of-transit orbits (and therefore mixes underlying trends and/or red noise together). The current data can be fit using a 2nd-order polynomial, or fit using a more physically motivated model including a sinusoidal component with a period equal to the planetary orbital period, representing the thermal phase variation due to the day-night temperature difference \citep{Knutson:2007bl}. Either model results in a better fit to the data than the linear slope for WASP-12 and WASP-19, and we decided to use the Bayesian Information Criterion (BIC; \citet{Schwarz:1978kf, Liddle:2004cz}) to determine whether the improvements from either of the more complex baseline models was sufficiently significant.  The BIC includes a strong penalty for including additional parameters, and therefore provides a robust technique to distinguish between models; $\Delta\mathrm{BIC}\ge2$ is considered to be positive evidence against the null hypothesis. The BIC was not increased using the non-linear baseline models for either target ($\Delta\mathrm{BIC}\sim-0.5$). However, the best-fit peak-to-trough amplitude of 0.0018$\pm$0.0006 for a possible sinusoidal component in the WASP-12 data is within the range predicted for the thermal phase variations of very hot planets \citep{Cowan:2011kw}, though it is smaller than the value measured using {\it Spitzer} \citep{Cowan:2012gp}. The best-fit amplitude for WASP-19 is similar to WASP-12 (0.0016$\pm$0.0007). We conclude that due to the low significance of the fit, the limited time sampling and ambiguities introduced by the divide-oot method, the nature of the curvature is highly uncertain and must therefore be investigated with more complete observations before conclusions as to its validity or physical nature can be made. The light curve for WASP-17 does not include a post-egress portion so we cannot evaluate the presence of a curved baseline. 

\begin{figure}[htb]
\centering
{
\includegraphics[width=85mm]{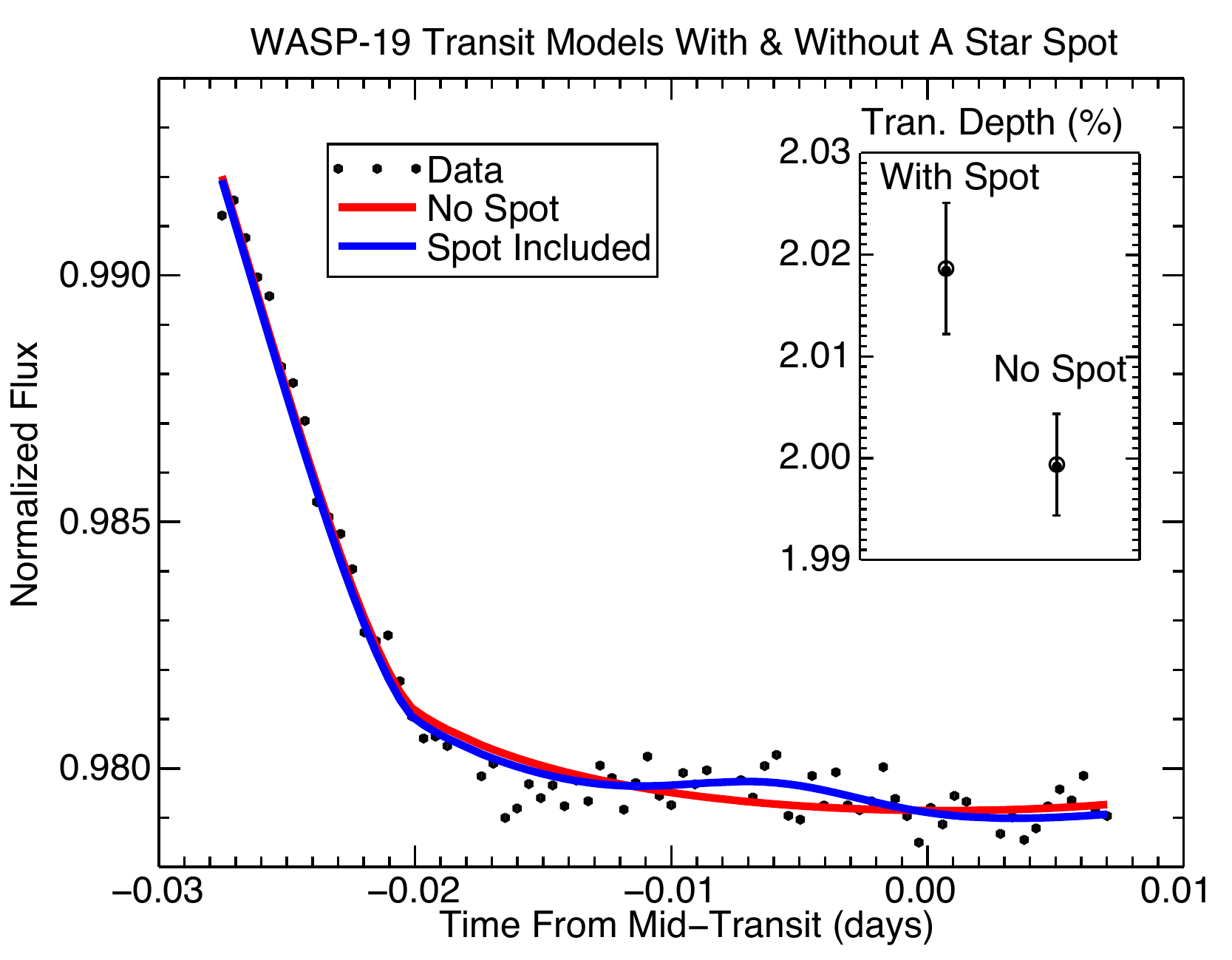}
 \caption{The trough of the transit for the band-integrated light curve for WASP-19, with models including standard transit model (red) and a model with a star spot (blue) over-plotted. The best-fit transit depths for each model are also plotted (inset). The value derived incorporating the spot model has a larger uncertainty from MCMC due to the additional free parameters, but the effects of red noise are not included and therefore the uncertainty on the spot-free fit is underestimated.}
 \label{fig:spot}
}
\end{figure}

The in-transit orbit of WASP-19 also has a region just after second contact (after the end of ingress) which deviates slightly from a standard transit curve (see Figure~\ref{fig:spot}). The amplitude and duration of the deviation is similar to the amplitude and duration of starspots detected in optical transit data by \citet{TregloanReed:2013gd}, so we experimented with including a Gaussian-shaped spot in our transit model. The spot model leads to a statistically better fit with $\Delta\mathrm{BIC}=7.8$ (see Figure~\ref{fig:spot}), leading us to adopt a model including a sunspot modeled as a Gaussian with a position centered at MJD 55743.526, a relative amplitude of 0.06\%, and a width of 0.0036 days. We locked the amplitude of the spot when fitting each of the bins, since our data quality is insufficient to determine variations with wavelength. Neither of our other data sets showed evidence for star spots, which is expected since both WASP-17 and WASP-12 are significantly hotter than WASP-19.  

Considering the ambiguity regarding the presence of additional visit-long components and star spots, we decided to use the average of all the model fits with and without a sinusoidal component or a spot for the band-integrated transit depth listed in Table~\ref{bfvals}, and augment the uncertainty values to encompass the full range of values; this increases the uncertainty by a factor of $\sim$4 for WASP-19 and a factor of $\sim$5 for WASP-12.  To remove these ambiguities in the band-integrated transit depth we would need a fully-sampled light curve and multiple visits to settle the question of spots; however, since we lock the values for any non-linear or spot components when fitting the bins, the final choice of the best-fit band-integrated model makes no difference in the relative depths for our wavelength bins.

\subsection{Fitting the Spectrally Binned Light Curves}
\label{BLCF}

Once we determined an adequate fit to the band-integrated light curves, we used the residuals of the fit to remove systematics common to all spectral channels (or bin of channels).  Our transit models for each individual channel include a constant scaling of these residuals, with the scale factor varying as a free parameter. This strategy is similar to methods developed independently by \citet{Deming:2012vd} and \citet{Stevenson:2013wf} (though without a scaling term for modulating the amplitude of the band-integrated residuals), and it obviates the need for using the divide-oot method. Additionally, we introduced two more components into the light curve model (each with a scaling factor as a free parameter) based on our measurements of the horizontal and vertical shifts of the spectrum on the detector over time. The scaling factors for these components are insignificant for most bins since a small shift for most points on the spectrum will not change the flux significantly; however, near spectral features or near the edges of the spectrum, these shifts can cause the flux within a single bin to drift up or down. Our final model light curve for comparison with the data takes the form 
\begin{multline}
\small LC_{final} = LC_{transit} * (a + bt + C_{1}*\mathrm{Res}_{BI} + \\ C_{2}*\mathrm{Shift}_{y} + C_{3}*\mathrm{Shift}_{x})
\end{multline}
where $LC_{transit}$ is the light curve model calculated using the \citet{Mandel:2002bb} prescription, $a$ and $b$ are coefficients for a linear trend with time, Res$_{BI}$ are the residuals from the band-integrated light curve, and the $C$ coefficients are scaling parameters determined through our MCMC fitting.

For the light curve for each spectral bin we followed the above methods for bad pixel and bad channel correction and then fit for the best model using MCMC. We locked the same parameters as with the band-integrated light curve, and additionally locked the limb darkening and mid transit time to the best-fit values from the band-integrated light curve analysis; this allows us to measure the relative change in transit depth while maintaining the same transit shape. We experimented with fitting for the limb darkening parameters using priors based on a linear interpolation between the J and H-band values from \citet{Claret:2011gy}, but we determined that there was no change in the final transit depths compared with exclusively using the band-integrated values.

\begin{figure}[htb]
\centering
{
\includegraphics[width=85mm]{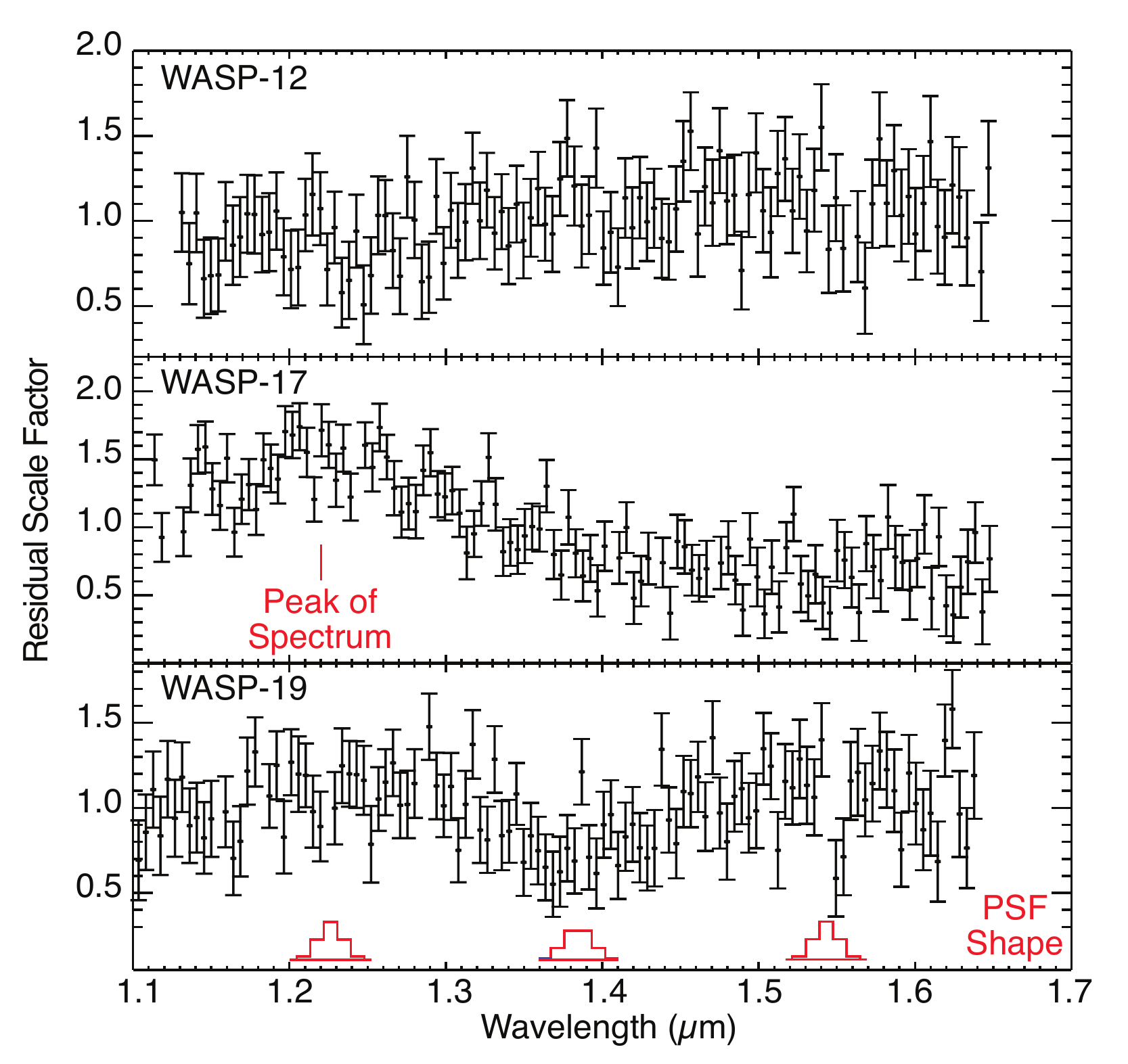}
 \caption{Best-fit scaling factors for the band-integrated light curve residuals derived for each channel (see \S\ref{BLCF}). The relative amplitude of the scaled-residuals component of the model changes with wavelength based on the peak illumination in each channel, and varies between targets based on the sub-array size and sampling mode (see Table~\ref{obs}).  For WASP-17 the scale factor peaks at the location of the peak flux in the spectrum, while for WASP-19 the scale factor varies based on the sampling of the spatial PSF.  WASP-12 has very little structure in the band-integrated residuals, and therefore shows no clear correlation with flux.}
 \label{fig:res_scaling}
}
\end{figure} 

\begin{figure}[htb]
\centering
{
\includegraphics[width=85mm]{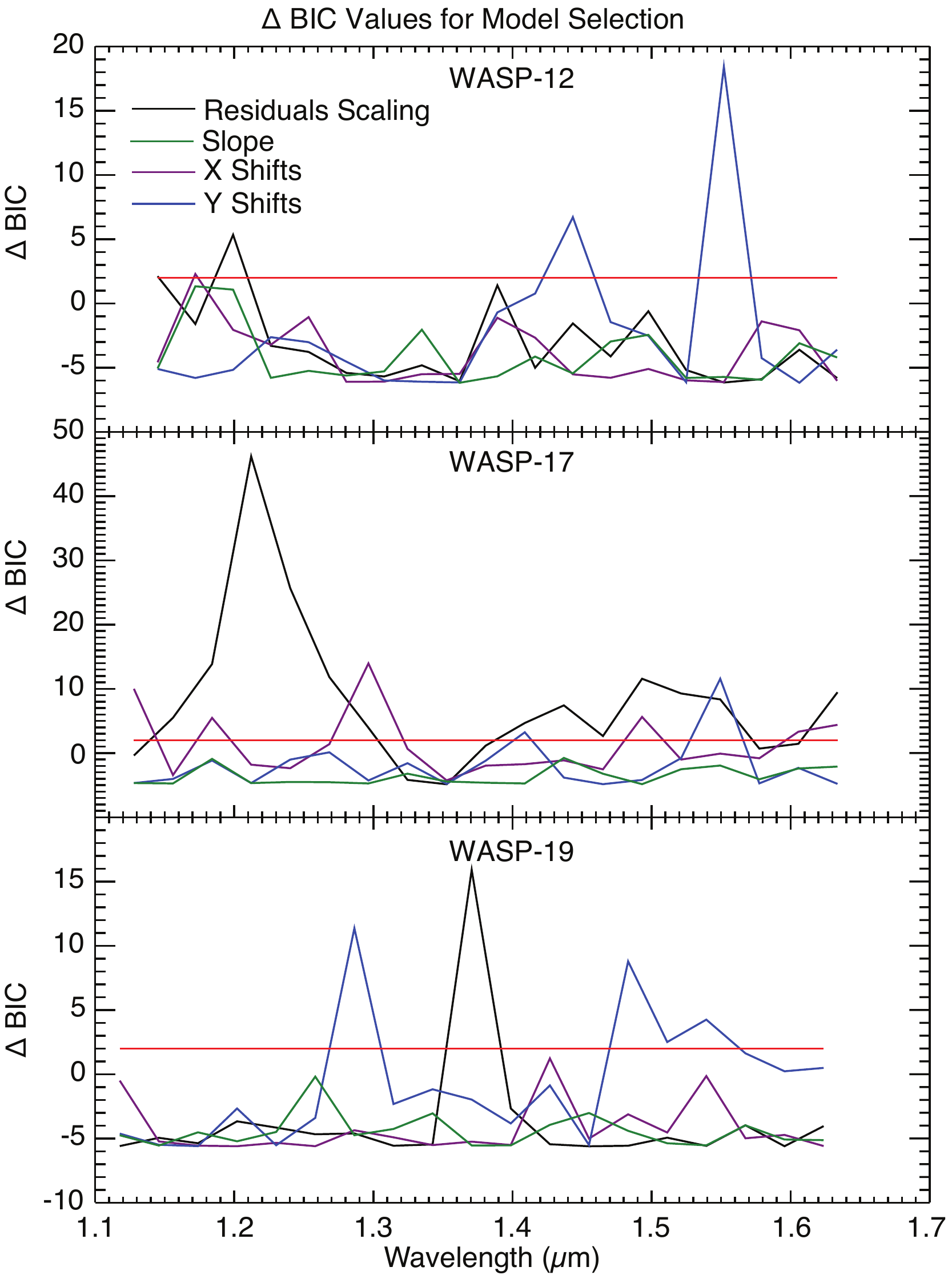}
 \caption{We calculate the change in BIC values for a model that fits for additional systematic trends (band-integrated residuals, the visit-long linear slope, x shift, y shift) compared with the default model (see \S\ref{BLCF}). $\Delta$BIC is shown for each of the 19 bins, for all targets (top: WASP-12, middle: WASP-17, bottom: WASP-19). The horizontal red line at zero indicates the level above which parameters are said to be significant --- parameters are only allowed to vary from the best-fit band-integrated values if they have $\Delta\mathrm{BIC}\ge2$.}
 \label{fig:BIC}
}
\end{figure}

\begin{figure}[htb]
\centering
{
\includegraphics[width=85mm]{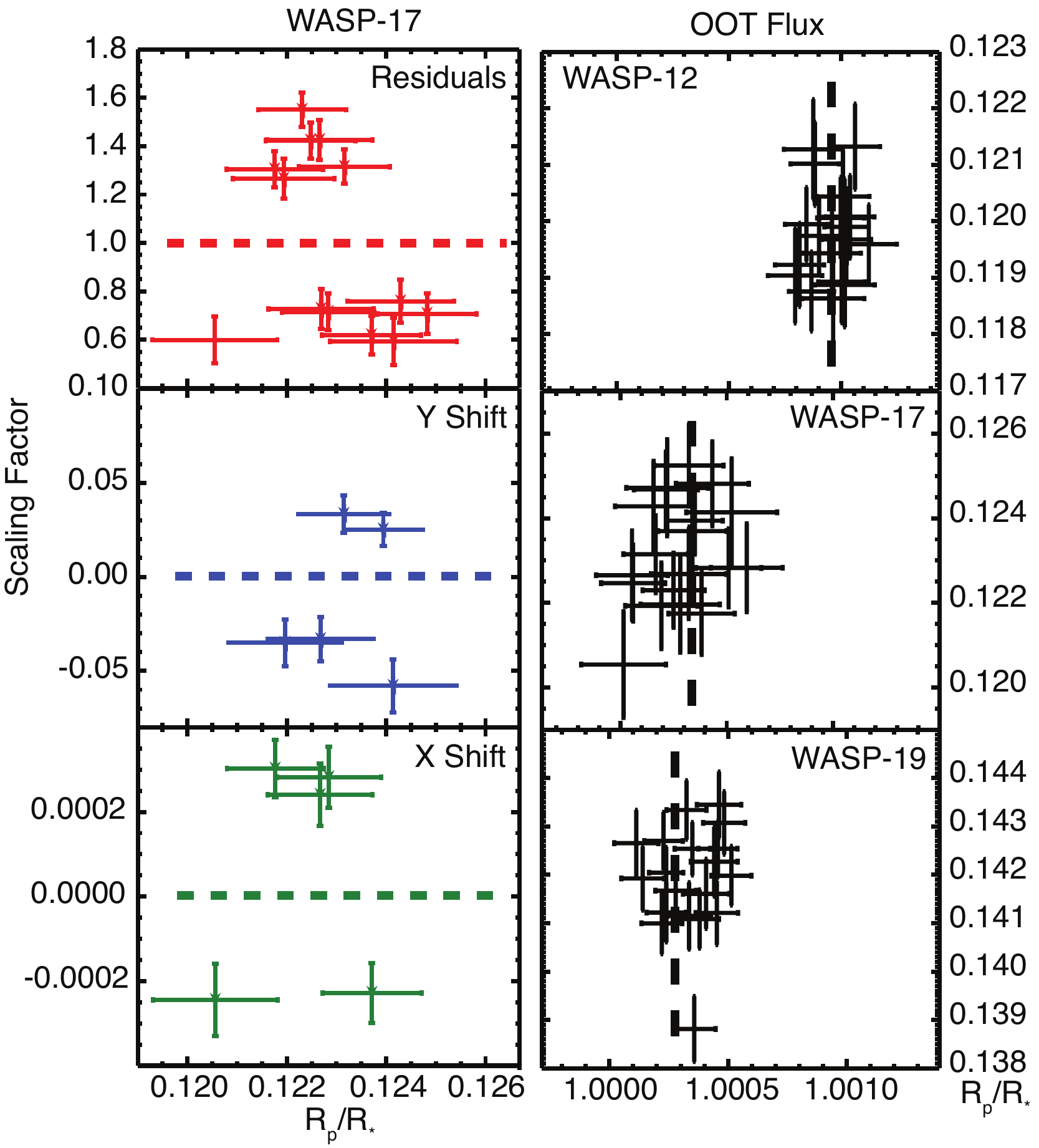}
 \caption{Left: Correlation plots for the three model components versus R$_{p}/$R$_{*}$, for WASP-17 (see \S\ref{BLCF}). Parameters were only allowed to vary for those bins in which doing so resulted in $\Delta\mathrm{BIC}\ge2$, and only the bins in which the parameters varied are plotted; the dotted lines represent the default value from the band-integrated results. Results for WASP-12 and WASP-19 are not plotted because the number of bins with open parameters for each component was small ($1-4$). Right: Best-fit out-of-transit flux versus R$_{p}/$R$_{*}$ for all the targets.  No correlation is seen between R$_{p}/$R$_{*}$ and any of the parameters.}
 \label{fig:correlations}
}
\end{figure}

In each bin the importance of the different systematic trends varies. The amplitude of the common-mode residuals is related to (but not directly correlated with) the peak intensity in each channel (see Figure~\ref{fig:res_scaling}), and the x shift is only important near spectral features or other steep gradients in the spectral direction. To avoid including unnecessary components in our light curve model, we examined the importance and validity of including each model parameter using a nested model selection analysis. We began by assuming that the values determined for the band-integrated light curve except for R$_{p}/$R$_{*}$ and the mean value of the out-of-transit flux would be valid for all the bins. We then calculated $\Delta$BIC for models with the inclusion of free parameters for the slope of the linear trend, the scale factor for the band-integrated residuals, and scale factors for components based on the x and y shifts; we included only the parameters that provided an improvement in the BIC ($\Delta\mathrm{BIC}\ge2$) over the model that locked that parameter. The $\Delta$BIC values for each of our 0.027\;$\mu$m-wide bins for each of our targets are shown in Figure~\ref{fig:BIC}.  To further confirm that we are not over-fitting our data, we searched for correlations between different free parameters in our light curve model and the final transit depths. Most of the parameters in most of the bins remain locked to the band-integrated values (the slope of the linear trend remained locked for every bin for all targets), and we see no evidence of correlations between parameters for the fitted parameter values in any of our targets (see Figure~\ref{fig:correlations}).

\begin{figure*}[htb]
\centering
{
\includegraphics[width=170mm]{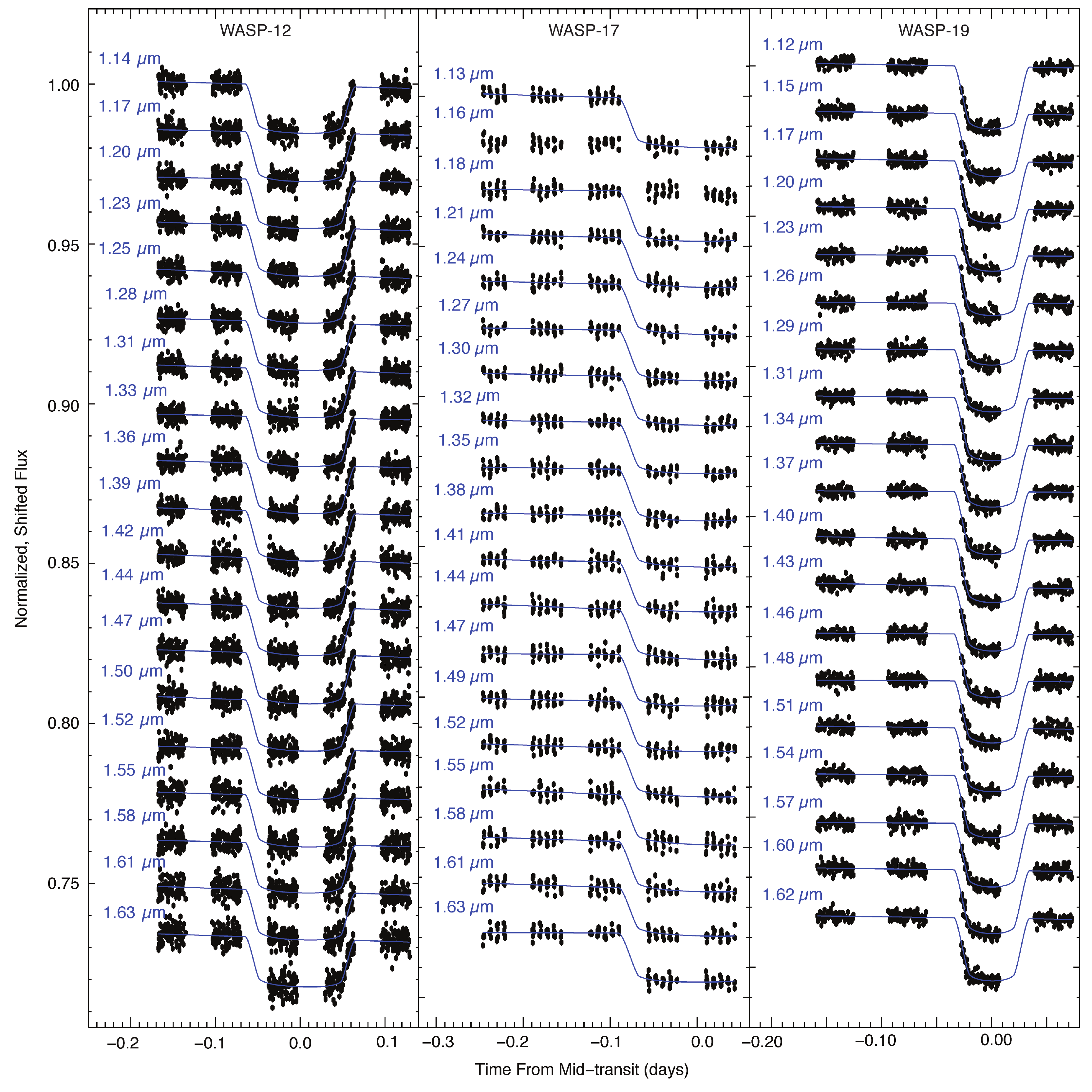}
\caption{The final results for all the bins for each target are shown in black, after removing time series components based on the scaled residuals from the band-integrated light curve, as well as any scaled components based on the spectral shift in the x and y directions that were deemed statistically significant (see \S\ref{BLCF}).  The best-fit transit model from our MCMC analysis is shown in blue. The light curves all show essentially white noise, with no evidence of correlated noise or remaining systematic trends.}
\label{fig:binfit}
}
\end{figure*}

\begin{figure*}[htb]
\centering
{
\includegraphics[width=170mm]{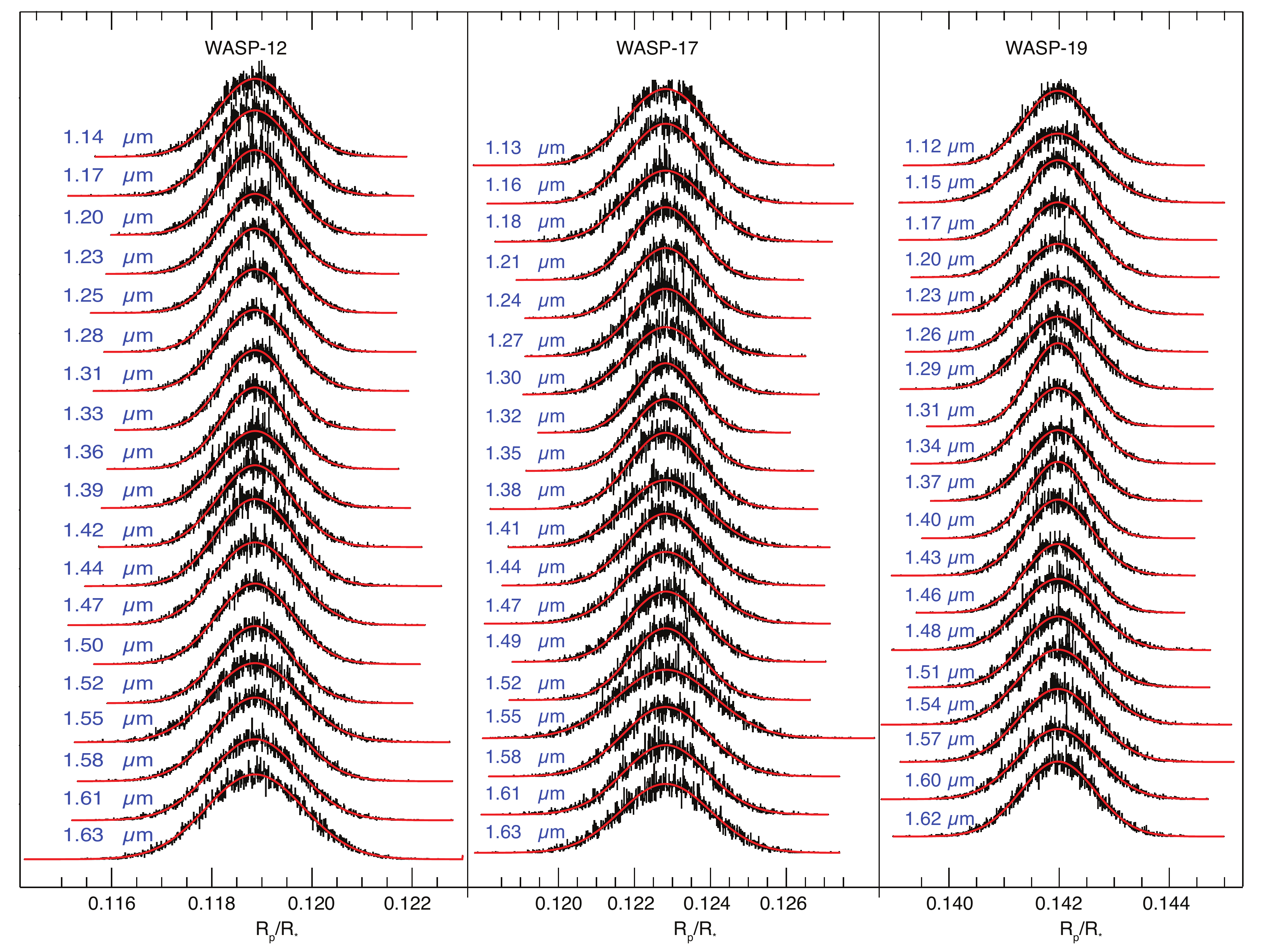}
 \caption{Posterior distributions from MCMC for R$_{p}/$R$_{*}$ for every bin, for each of the three targets.  All of the final distributions are symmetric and well-approximated by a Gaussian fit (red).}
 \label{fig:posteriors}
}
\end{figure*}

\begin{figure}[htb]
\centering
{
\includegraphics[width=85mm]{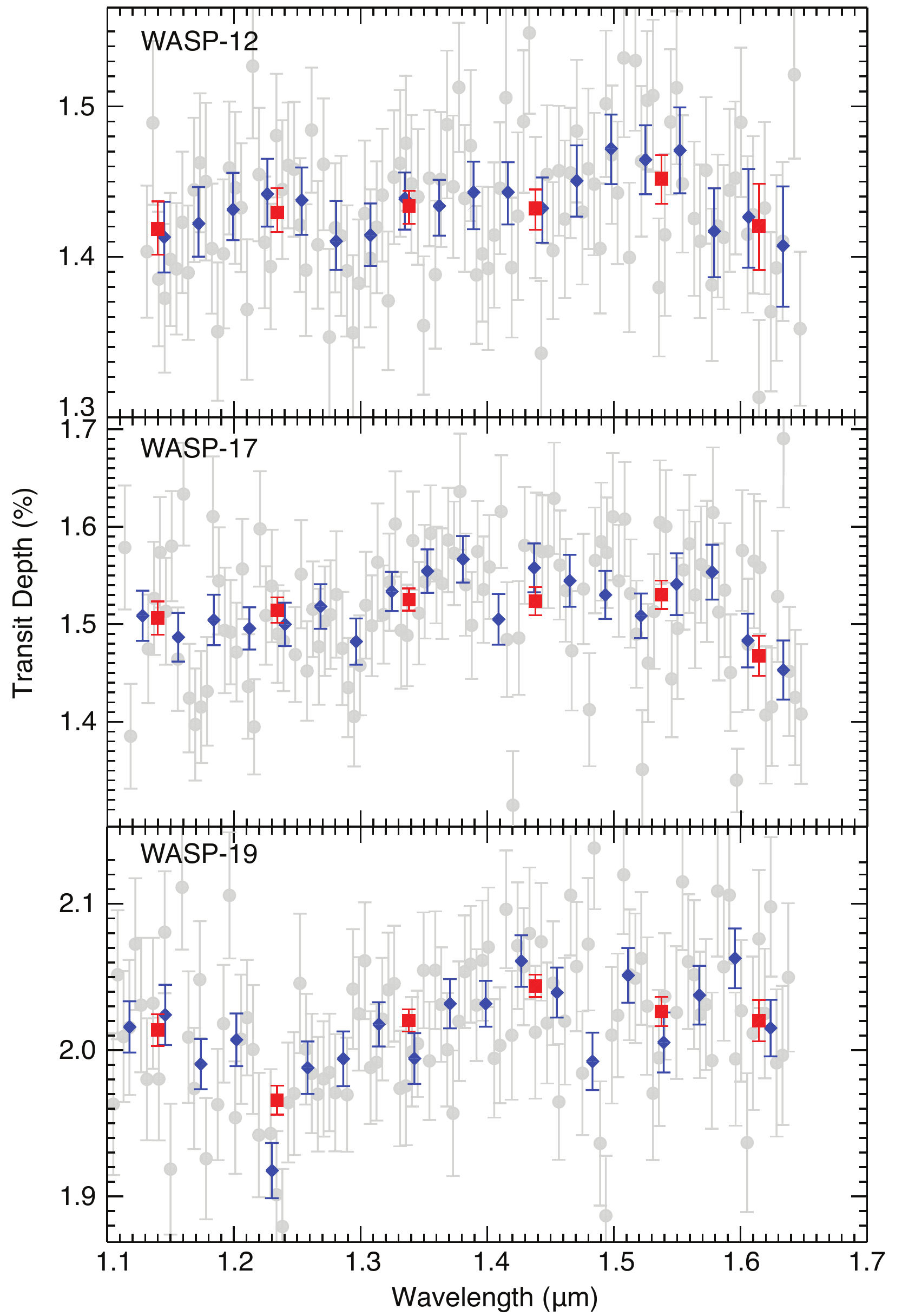}
 \caption{Final spectra for each of our targets.  The individual channel depths are shown in grey, with the results for 0.027\;$\mu$m-wide (blue) and 0.1\;$\mu$m-wide (red) bins overplotted. The differences between the channels and the 0.027\;$\mu$m-wide bins are consistent with photon-noise variations, but the 0.1\;$\mu$m-wide bins appear to remove structure in the spectra that could be significant; we therefore chose to use the 0.027\;$\mu$m-wide bins in our analysis.}
\label{fig:finalspec}
}
\end{figure}

\begin{figure}[htb]
\centering
{
\includegraphics[width=85mm]{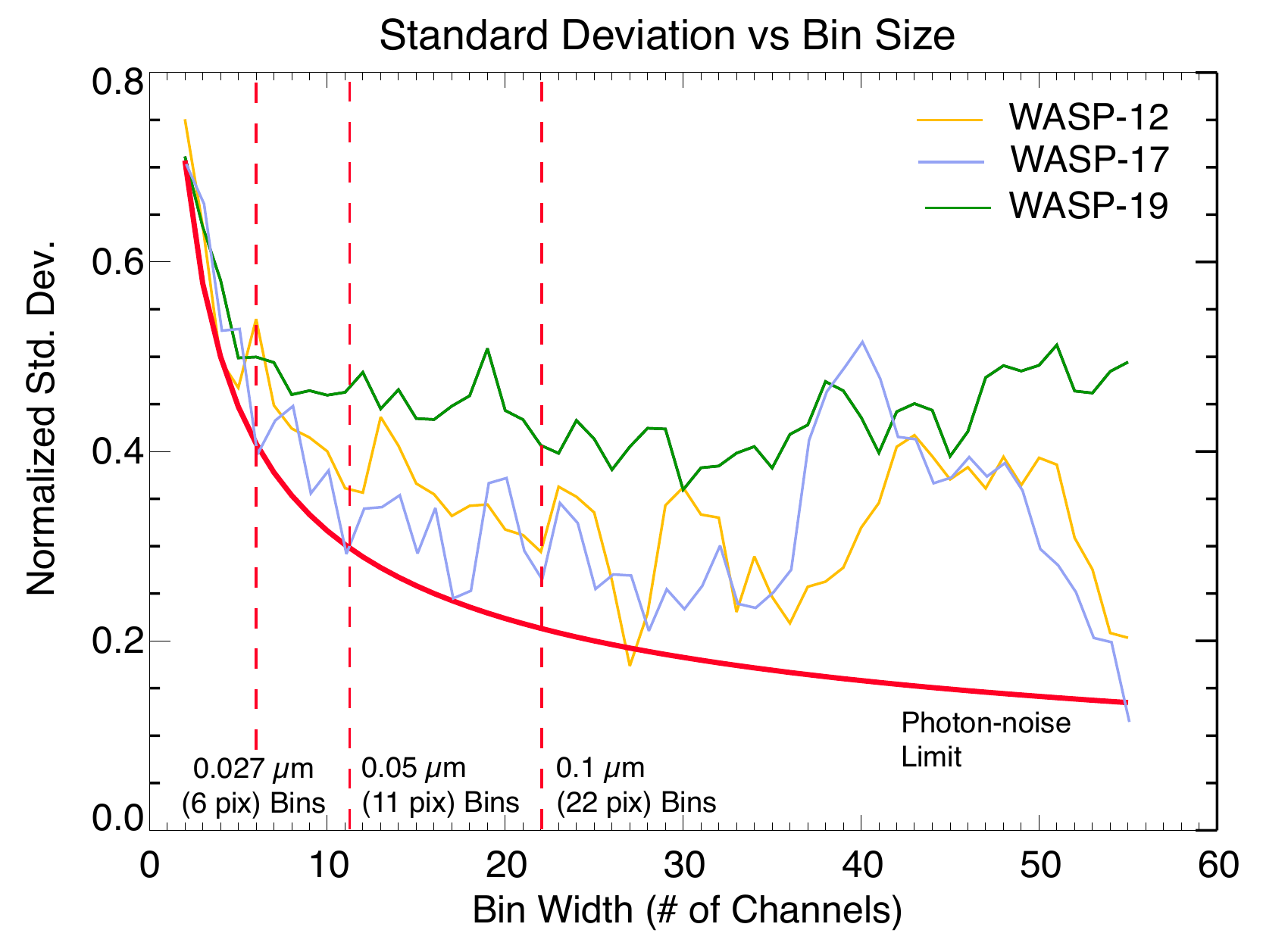}
 \caption{After fitting for the transit depths using individual channels, we binned the spectra using bin sizes between 2 and 55 points and then calculated the standard deviation of each binned spectrum; the results for each star are plotted as well as the expected relationship based on photon-noise statistics alone.  The standard deviation for all the targets is approximately photon-limited up to 5-channel bins, but then levels off.  We use a 6-channel bin size for our final results; spectra using both the 6-channel bins and 22-channel (0.1\;$\mu$m) bins are shown in Figure~\ref{fig:finalspec} for comparison.}
 \label{fig:photon_noise_test}
}
\end{figure}

In Figure~\ref{fig:binfit} we show final light curves for all of our 0.027\;$\mu$m-wide bins for each target after the various best-fit systematic trend components have been removed; they are overplotted with the best-fit transit light curve model.  The light curves show no sign of correlated noise, and the posterior distributions (shown in Figure~\ref{fig:posteriors}) are all fit well by a Gaussian distribution.  Our final spectra for each of our science targets are shown in Figure~\ref{fig:finalspec}; we plot the best-fit transit depth values for each individual channel, and two bin sizes (0.027\;$\mu$m and 0.1\;$\mu$m).  The individual channels clearly show a high point-to-point scatter which appears to be largely due to photon noise, so we experimented with binning the channels using sequential bin sizes (2 channels, 3 channels, etc).  The rms of the resulting spectra drops off quickly, but then stays elevated above the photon-noise limit for all stars beyond a 5-channel bin width, suggesting structure in the spectrum on scales larger than 5 pixels (see Figure~\ref{fig:photon_noise_test}).  We therefore chose to use the 6-channel bins (0.027\;$\mu$m) for our final spectrum, since they will largely conserve the overall structure of the individual-channel spectrum while decreasing the photon noise considerably and allowing for improved removal of systematic trends.    Larger bin sizes, as used by \citet{Stevenson:2013wf} and \citep{Huitson:2013ud}, do not fully encapsulate the structure in the smaller-bin spectrum. This smoothing is not incorporated into the uncertainty limits for the wider bins since the uncertainty is purely based on the goodness-of-fit of the transit model; we therefore believe the use of bin sizes $<0.03\;\mu$m is necessary to avoid misinterpretation of spectral characteristics.  The best-fit transit depths for the 0.027\;$\mu$m-wide bins for all of our targets are listed in Table~\ref{data}.

\begin{deluxetable*}{cc | cc | cc}
\tablecaption{Derived Transit Depths For Binned Data}
\tabletypesize{\scriptsize}
\tablewidth{0pt}
\tablehead{ \multicolumn{2}{c}{WASP-12 b}  & \multicolumn{2}{c}{WASP-17 b}  & \multicolumn{2}{c}{WASP-19 b}  \\ \multicolumn{1}{c}{$\lambda$ ($\mu$m)} &  \multicolumn{1}{c}{Transit Depth (\%)} &\multicolumn{1}{c}{$\lambda$ ($\mu$m)} & \multicolumn{1}{c}{Transit Depth (\%)} & \multicolumn{1}{c}{$\lambda$ ($\mu$m)} & \multicolumn{1}{c}{Transit Depth (\%)}}

\startdata
1.145 & 1.4131 $\pm$ 0.0235 & 1.128 & 1.5087 $\pm$ 0.0257 & 1.118  & 2.0159 $\pm$ 0.0175 \\
1.172 & 1.4211 $\pm$ 0.0232 & 1.156 & 1.4867 $\pm$ 0.0250 & 1.146  & 2.0241 $\pm$ 0.0206 \\
1.199 & 1.4302 $\pm$ 0.0224 & 1.184 & 1.5044 $\pm$ 0.0259 & 1.174  & 1.9905 $\pm$ 0.0172 \\
1.226 & 1.4417 $\pm$ 0.0226 & 1.212 & 1.4957 $\pm$ 0.0216 & 1.202  & 2.0071 $\pm$ 0.0180 \\
1.253 & 1.4376 $\pm$ 0.0224 & 1.240 & 1.4998 $\pm$ 0.0222 & 1.230  & 1.9269 $\pm$ 0.0189 \\
1.281 & 1.4103 $\pm$ 0.0230 & 1.268 & 1.5166 $\pm$ 0.0226 & 1.258  & 1.9880 $\pm$ 0.0180 \\
1.308 & 1.4143 $\pm$ 0.0207 & 1.296 & 1.4822 $\pm$ 0.0237 & 1.286  & 1.9941 $\pm$ 0.0187 \\
1.335 & 1.4387 $\pm$ 0.0190 & 1.325 & 1.5362 $\pm$ 0.0197 & 1.314  & 2.0176 $\pm$ 0.0151 \\
1.362 & 1.4338 $\pm$ 0.0186 & 1.353 & 1.5545 $\pm$ 0.0223 & 1.343  & 1.9943 $\pm$ 0.0174 \\
1.389 & 1.4419 $\pm$ 0.0225 & 1.381 & 1.5686 $\pm$ 0.0239 & 1.371  & 2.0318 $\pm$ 0.0168 \\
1.416 & 1.4414 $\pm$ 0.0207 & 1.409 & 1.5050 $\pm$ 0.0261 & 1.399  & 2.0317 $\pm$ 0.0157 \\
1.443 & 1.4322 $\pm$ 0.0217 & 1.437 & 1.5578 $\pm$ 0.0250 & 1.427  & 2.0546 $\pm$ 0.0176 \\
1.471 & 1.4505 $\pm$ 0.0237 & 1.465 & 1.5446 $\pm$ 0.0267 & 1.455  & 2.0363 $\pm$ 0.0171 \\
1.498 & 1.4719 $\pm$ 0.0231 & 1.493 & 1.5300 $\pm$ 0.0247 & 1.483  & 1.9923 $\pm$ 0.0196 \\
1.524 & 1.4645 $\pm$ 0.0229 & 1.521 & 1.5086 $\pm$ 0.0229 & 1.511  & 2.0470 $\pm$ 0.0187 \\
1.552 & 1.4707 $\pm$ 0.0286 & 1.549 & 1.5410 $\pm$ 0.0316 & 1.539  & 2.0053 $\pm$ 0.0205 \\
1.579 & 1.4170 $\pm$ 0.0296 & 1.577 & 1.5534 $\pm$ 0.0282 & 1.568  & 2.0350 $\pm$ 0.0196 \\
1.606 & 1.4264 $\pm$ 0.0329 & 1.606 & 1.4875 $\pm$ 0.0278 & 1.597  & 2.0578 $\pm$ 0.0197 \\
1.633 & 1.4073 $\pm$ 0.0400 & 1.634 & 1.4530 $\pm$ 0.0303 & 1.624  & 2.0142 $\pm$ 0.0188 \\
\enddata
\label{data}
\end{deluxetable*}

\subsection{Uncertainty Analysis}
\label{EA}

The uncertainty limits for our light curve parameters were derived from the widths of our MCMC posterior probability distributions; however, the uneven sampling before and after a transit as well as across a transit event due to the gaps in the HST orbit make the calculation of the expected noise limit difficult. We therefore decided to construct synthetic data sets for each of our targets in order to identify the different contributing sources of uncertainty in the final results, with each synthetic data set for an exoplanet constructed using the best-fit parameters from the fit to our band-integrated light curve and the timing array of our real data. Stochastic Gaussian noise was injected at the level of the final rms determined for our data, and the synthetic data was fit using MCMC in the same method described above for the real light curves. Since each data set has a relatively small number of data points (131 for WASP-17, 274 for WASP-19, and 484 for WASP-12), the impact of outliers due to purely stochastic noise can have a considerable effect, so we repeated this process 100 times with different randomly generated noise distributions in order to determine the range of uncertainties produced by MCMC. We can then compare the predicted noise based on the number of points in transit to the predicted uncertainty from MCMC fits to the synthetic data to estimate the increase in uncertainty due to the uneven sampling of the light curves. Also, by comparing the uncertainty derived for our real data to the range of uncertainties for the simulated data sets we can estimate the amount of additional (red) noise in our data compared with a purely (white) stochastic noise distribution.

\begin{deluxetable}{lccc}
\tablecaption{Uncertainty Analysis}
\tabletypesize{\scriptsize}
\tablewidth{85mm}
\tablehead{\colhead{Parameters} &
                    \colhead{WASP-12} &
                    \colhead{WASP-17} &    
                    \colhead{WASP-19}}
\startdata
Data points during transit & 196 & 54 & 70 \\
Data points out of transit & 288 & 77 & 204 \\
\cutinhead{Band Integrated Time Series} 
Photon noise (ppm) & 357 & 279 & 255 \\
RMS of residuals (ppm) & 515 & 350 & 305 \\
\vspace{-1mm} \\
Predicted\tablenotemark{1} $\sigma_{td}$  (ppm)& 52 & 67 & 45 \\
$\sigma_{td}$ from MCMC, Data & 53 & 144 & 65 \\
$\sigma_{td}$ from MCMC, Sim.\tablenotemark{2} & 53$\pm$2 & 145$\pm$13 & 63$\pm$11 \\
\vspace{-1mm} \\
RMS/photon noise & 1.44 & 1.26 & 1.20 \\
Data/Pred. & 1.02 & 2.15 & 1.44 \\
Sim./Pred. & 1.03$\pm$0.04 & 2.16$\pm$0.19 & 1.40$\pm$0.24 \\
\cutinhead{0.027\;$\mu$m Bin Width (19 Total)} 
Photon noise (ppm) & 1560 & 1220 & 1110 \\
RMS of residuals (ppm) & 1880 & 1400 & 1230 \\
\vspace{-1mm} \\
Predicted\tablenotemark{1} $\sigma_{td}$ (ppm) & 174 & 249 & 170 \\
$\sigma_{td}$ from MCMC, Data  & 180 & 257 & 180 \\
$\sigma_{td}$ from MCMC, Sim.\tablenotemark{2} & 181$\pm$6 & 242$\pm$14 & 187$\pm$11 \\
\vspace{-1mm} \\
RMS/photon noise & 1.22 & 1.15 & 1.11 \\
Data/Pred.  & 1.02 & 1.03 & 1.06 \\
Sim./Pred. & 1.03$\pm$0.03 & 0.97$\pm$0.06 & 1.1$\pm$0.06 \\

\enddata
\tablenotetext{1}{Calculated from the residual rms and the number of points during transit and out of transit}
\tablenotetext{2}{Simulated data was created with a sampling equivalent to that of the real data, and an rms equivalent to the rms of the final residuals.}
\label{whlerror}
\end{deluxetable}

We also explored the use of residual-permutation analysis (RP) to estimate the effects of red noise. We fit the light curves using Levenberg--Marquardt least-squares fitting, subtracted the best fit model from the light curve, shifted the residuals by one position and then added the model back in and re-fit the data, cycling through all the data points in each light curve. However, we found that with such a small number of data points in our light curves and the uneven sampling of the HST orbits the RP method is not sufficiently robust; the final distributions for the fitted values of $R_{p}/R_{*}$ showed a large scatter without any clear pattern. We therefore relied on our simulated data tests to determine how close we came to the expected photon noise.

The band-integrated photon noise statistics, rms uncertainty, and uncertainties in transit depth determined from MCMC fitting for the real and synthetic data sets are shown in Table~\ref{whlerror}. We find that the rms of the data is $1.2-1.44\times$ the expected photon noise for band-integrated time series, but only $1.11-1.22\times$ the photon noise limit for the binned data. For WASP-12 the MCMC results for the synthetic data match within a few percent to the predicted uncertainties based on the rms, suggesting that the impact of light curve sampling is minimal. The real band-integrated data for WASP-12 are slightly noisier than the synthetic data suggesting some correlated noise, most likely due to trends in the out-of-transit portion of the data discussed previously \S\ref{W19anom}. The WASP-19 results are similar, though the MCMC uncertainties and the dispersion in the range of value for the synthetic data are larger than predicted due to the impact of fitting for the presence of a spot (\S\ref{W19anom}). For WASP-17 the uncertainty for the synthetic data is more than $2\times$ larger than the predicted uncertainty due to the lack of data covering ingress/egress or post-transit. However, we note that the effects of sampling and correlated noise are almost completely neutralized in the binned data by our residual subtraction - the ratio of the uncertainty for the simulated data to the analytical prediction for all the targets drops to essentially unity, demonstrating the effectiveness of our component removal method.

\section{Discussion}
\label{disc}

The observations analyzed in this study represent a preliminary sample of hot exoplanets observed with the WFC3 instrument on HST.  The three planets include two extremely hot planets with temperature structures constrained by {\it Spitzer} occultation data (WASP-12 b and WASP-19 b) as well as a cooler planet with a highly-inflated planetary radius (WASP-17 b), allowing us to investigate two classes of planets that pose significant challenges for current theories of exoplanet structure and evolution.

\subsection{Comparison with Atmospheric Models}
\label{AC}

Absorption band depths in transit spectra probe the line of sight through the terminator of the planet, and are primarily sensitive to a combination of the atmospheric composition and the scale heights over which each species is absorbing. These factors can be significantly degenerate and it is difficult to place strong constraints on the overall abundances of different species with observations in only a single wavelength band.  We therefore reserve a detailed examination of constraints on atmospheric composition and structure to a later study, and restrict our current analysis to a discussion of the general implications of qualitative comparison with several different sets of models.

\begin{figure*}[htb]
\centering
{
\includegraphics[width=170mm]{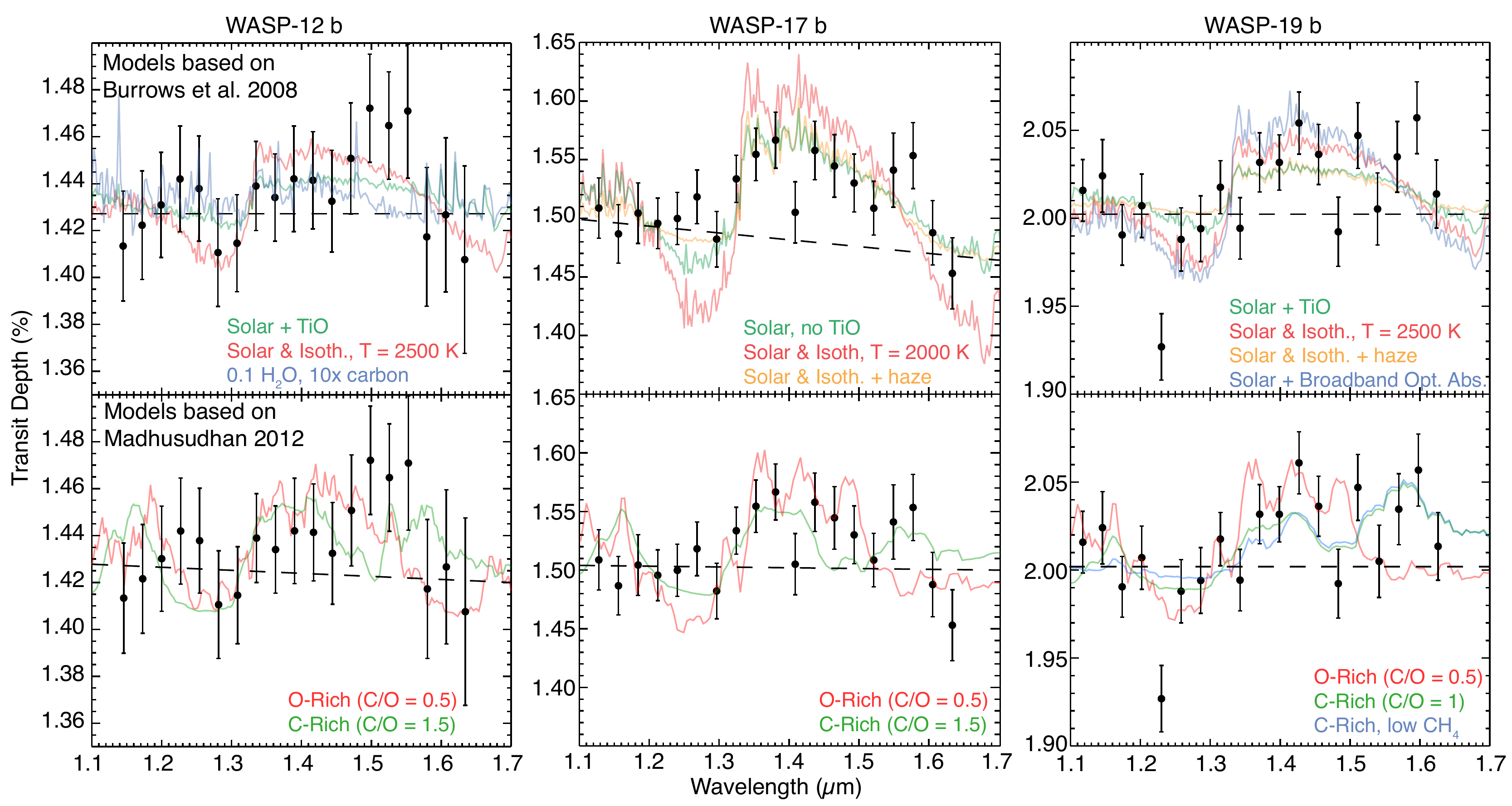}
\caption{Transit depths for each of the 19 bins for each target, with models based on the framework of Burrows et al. (top) and Madhusudhan et al. (bottom). Standard models from Burrows et al. provide a good fit for WASP-17 b and a reasonable fit for WASP-12 b, but for WASP-19 b the models do not fit well beyond 1.45\;$\mu$m. Models with a deep water absorption feature can also be adjusted to fit the data by adding an absorbing haze layer with an opacity of 0.01 cm$^2$/g; the hazy model for WASP-17 b is further supported by the linear slope that is needed to match the models to the data. The oxygen-rich and carbon-rich models by Madhusudhan et al. fit equally well for WASP-12 band WASP-17 b, but for WASP-19 b the carbon-rich models  provide a statistically better fit than the oxygen-rich models. However, except for WASP-17 b the data is fit almost equally well by a flat spectrum, though WASP-19 b would require a very large scatter between the data points.}
 \label{fig:atmmodels}
}
\end{figure*}

In Figure~\ref{fig:atmmodels} we plot the data for each planet and overplot two different sets of models, which utilize different strategies for constraining the atmospheric structure and composition. One set (top in Figure~\ref{fig:atmmodels}) is based on the framework of \citet{Burrows:2000ho} and more recently \citet{Burrows:2006ga}, \citet{Burrows:2008ff} and \citet{Howe:2012et}. The Burrows models calculate the chemical and radiative equilibrium state of each planet based on the mass, size, and incident radiation, assuming solar abundances; the spectra were then calculated by combining day- and night-side model atmospheres joined at the terminator. Adjustments were made to the abundance of important molecular absorbers such as H$_2$O, CH$_4$ and CO and/or the inclusion of additional absorbers that affect the temperature structure and/or broadband optical depth of the atmosphere with the goal of improving fits to multi-wavelength observations. For example, additional opacity at optical wavelengths is required to produce a thermal inversion postulated to explain {\it Spitzer}/IRAC photometric measurements during occultation for a number of planets including WASP-12 b \citep{Cowan:2012gp, Crossfield:2012gm} and possibly WASP-19 b \citep{Anderson:2013dx}. TiO has been considered as the most likely candidate \citep{Hubeny:2003eb, Fortney:2008ce}, but the lifetime for TiO in the upper atmosphere may be problematic for this hypothesis \citep{Spiegel:2009ja} and recent searches for spectral features of TiO have been unsuccessful \citep{Huitson:2013ud}. On the other hand, a haze or dust with opacity through the optical and NIR is required to fit measurements of molecular absorption features for several hot Jupiters \citep{Charbonneau:2002er, Pont:2013ch, Deming:2013vu}. While the physical nature of these absorbers is currently unclear, we can test how different opacities for these parameters affect the model spectra in our wavelength region.

The Burrows models, which are characterized by broad H$_2$O absorption at 1.4\,$\mu$m that slopes consistently downward towards longer wavelengths, fit the data for WASP-17 b reasonably well --- both a standard model and an isothermal model with haze yield a lower BIC (assuming 3 degrees of freedom) than simply fitting a line to the data (2 degrees of freedom), with the best-fitting model (the hazy model) giving a $\Delta\chi^2 \sim10$. A model with haze is required to reproduce the flat region shortwards of 1.3\;$\mu$m, and a haze hypothesis may gain additional support from the fact that the best fits to the models are improved ($\Delta\chi^2 <0$) in every case by including a linear trend to the models; we discuss the implications of these results in \S\ref{comp}. However, the results for the two hotter planets are more ambiguous.  The majority of the spectrum for WASP-12 b is consistent with a flat spectrum within the uncertainties ($\chi^{2}_{red} = 0.57$), and the amplitude of the expected features do not allow us to discriminate between standard models with either an equilibrium temperature structure, an isothermal temperature structure suggested by \citet{Crossfield:2012gm}, or a model with a deficit of water and enhanced carbon abundance that best fits the analysis of {\it Spitzer}/IRAC occultation results by \citet{Cowan:2012gp}. WASP-12 b and possibly WASP-17 b also appear to have additional absorption in the region from $1.5-1.6\;\mu$m; these features are several bins wide, and do not appear to be the result of random noise. For WASP-19 b the results are even less consistent with the models - none of the models yield an improvement in BIC or $\chi^2$ over a linear fit.  The spectrum shows an increase in absorption beyond 1.35\;$\mu$m suggestive of H$_2$O but does not include the consistent drop at longer wavelengths expected from the models and apparent in the WASP-17 b spectrum; additionally, several bins in this region show a steep drop in absorption compared with the smooth downward trend expected from the Burrows models.

The second set of models we compare to our data (bottom in Figure~\ref{fig:atmmodels}) are based on the framework of \citet{Madhusudhan:2009gd} and \citet{Madhusudhan:2012ga}, which relax the stringent requirements for radiative and chemical equilibrium in favor of flexibility when exploring the constraints on parameter space from available observations. In particular, the Madhusudhan models explore a range of carbon-to-oxygen (C/O) ratios for the overall composition of the atmosphere, and include a number of less abundant carbon-bearing species that may produce additional absorption features in NIR spectra at C/O$\ge1$.  The models plotted roughly correspond to either an oxygen-rich chemistry (C/O $\sim0.5$, i.e. essentially the solar value) or a carbon-rich chemistry (C/O $\gsim1$) for specific temperature profiles (see \citet{Madhusudhan:2012ga} for details). It is clear that there are a number of overlapping spectral features that lead to degeneracies - the H$_2$O feature at 1.4\,$\mu$m overlaps with CH$_4$ at 1.36\,$\mu$m and HCN at 1.42\,$\mu$m-1.51\,$\mu$m, while the H$_2$O feature at 1.15\,$\mu$m overlaps with CH$_4$. The oxygen-rich and carbon-rich models primarily diverge between 1.45 and 1.65\;$\mu$m, where the carbon-rich models include features from HCN and C$_2$H$_2$; while the additional absorption in WASP-17 b and WASP-19 b appears to line up well with these features and produces an improvement in $\chi^2$, the uncertainties in both our data and the range of potential model parameter values are large enough that we cannot discriminate between oxygen-rich and carbon-rich compositions based on these data alone.

We conclude that the data for all our targets are consistent for the most part with standard atmospheric models, but further improvements in S/N and a more comprehensive modeling strategy incorporating additional constraints on the molecular abundances and temperatures from other data sets are necessary to discriminate between them. In particular, the origin of significant deviations from the standard solar composition model predictions at wavelengths beyond 1.5\;$\mu$m is unclear; these features could either be indicative of unexpected atmospheric absorption features or they could be unexplained artifacts in the data. We have examined all of our data analysis routines in detail and we have found no obvious problems with the analysis of these bins, but repeated observations are necessary to confirm that the results are robust. We also point out the importance of using bins appropriately sized to be sensitive to the possibility of narrower spectral features in the data; Figure~\ref{fig:finalspec} demonstrates that using bin sizes larger than $\sim0.03\;\mu$m smoothes the data significantly and has the potential to erase the signatures of small-scale fluctuations in the data.

\subsection{Comparison to Previous Results}
\label{comp}

\subsubsection{WASP-12}
\begin{figure}[htb]
\centering
{
\includegraphics[width=85mm]{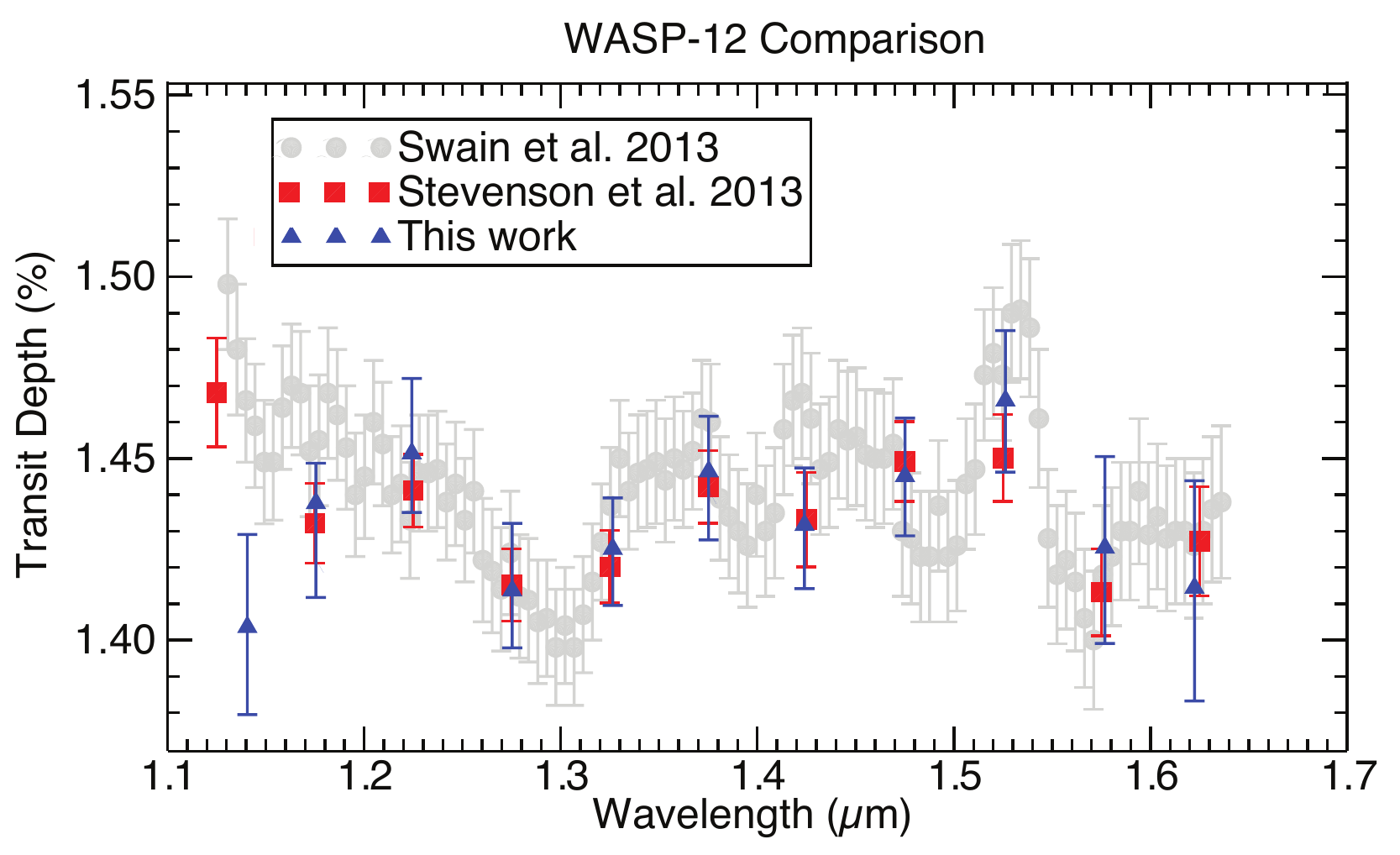}
\caption{Results from \citet{Swain:2013hn} shown in grey, results from \citet{Stevenson:2013wf} shown in red, and from this work in blue. Results from this work have been binned to the same size and number of bins as those used by \citet{Stevenson:2013wf}, with edge bins offset due to different choices of spectral trimming. Results from \citet{Stevenson:2013wf} have been shifted up slightly for comparison.  The spectra are largely consistent, with the most noticeable offsets visible at the short edge of the spectrum. }
 \label{fig:swain}
}
\end{figure}

As mentioned previously, the data set that we analyzed for WASP-12 was originally observed and analyzed by \citet{Swain:2013hn}, and the data set has also recently been analyzed as part of a multi-wavelength study by \citet{Stevenson:2013wf}. Figure~\ref{fig:swain} shows our final spectrum for WASP-12 binned to match \citet{Stevenson:2013wf} and plotted with the results from these two studies. While it is always difficult to pin-point differences between independent analyses, there are two possible sources of significant variations between the results of the three different studies: the technique for fitting or modeling the flux from the nearby contaminating source, and the details of fitting the transit light curve model.  \citet{Stevenson:2013wf} demonstrated that by using two different transit modeling methods, small differences could be introduced in the spectrum; similarly, we have shown in \S\ref{FFBC} that the choice of the spectrum for the contaminating flux from WASP-12 BC can change the fitted transit depths by a factor comparable to the fitting uncertainty. 

Remarkably, all the spectra show similar trends at wavelength longer than 1.2\,$\mu$m, with a high point at 1.225\,$\mu$m and a broad peak from $1.325 - 1.575\;\mu$m.  There are slight differences (at the $1-2\sigma$ level) for the bins at 1.425 and 1.525\;$\mu$m, but the major disagreement is at the short-wavelength edge of the spectrum - the Swain et al. results show a steady rise at short wavelengths while the Stevenson et al. results show a upward spike in the shortest-wavelength bin (1.125\;$\mu$m), in contrast our spectrum which shows a drop shortwards of 1.2\;$\mu$m.  This region of the spectrum is particularly susceptible to the choice of the dilution factor for the contaminating star due to the wavelength shift of the spectrum (see Figure~\ref{fig:contam_compare}), and the edges of the spectrum also exhibit a steep gradient in flux due to the grism sensitivity which can lead to systematic trends if the spectrum drifts over time (see \S\ref{Sys}); we therefore believe that a careful treatment of this spectral region is imperative. The downward slope of our final spectrum does not require any additional absorption from species such as TiH or CrH, as suggested by \citet{Swain:2013hn}.  Our uncertainties are larger than those of \citet{Stevenson:2013wf}, but we believe the larger uncertainties are warranted based on the uncertainty in the contribution from WASP-12 BC.

\subsubsection{WASP-17}
\begin{figure}[htb]
\centering
{
\includegraphics[width=85mm]{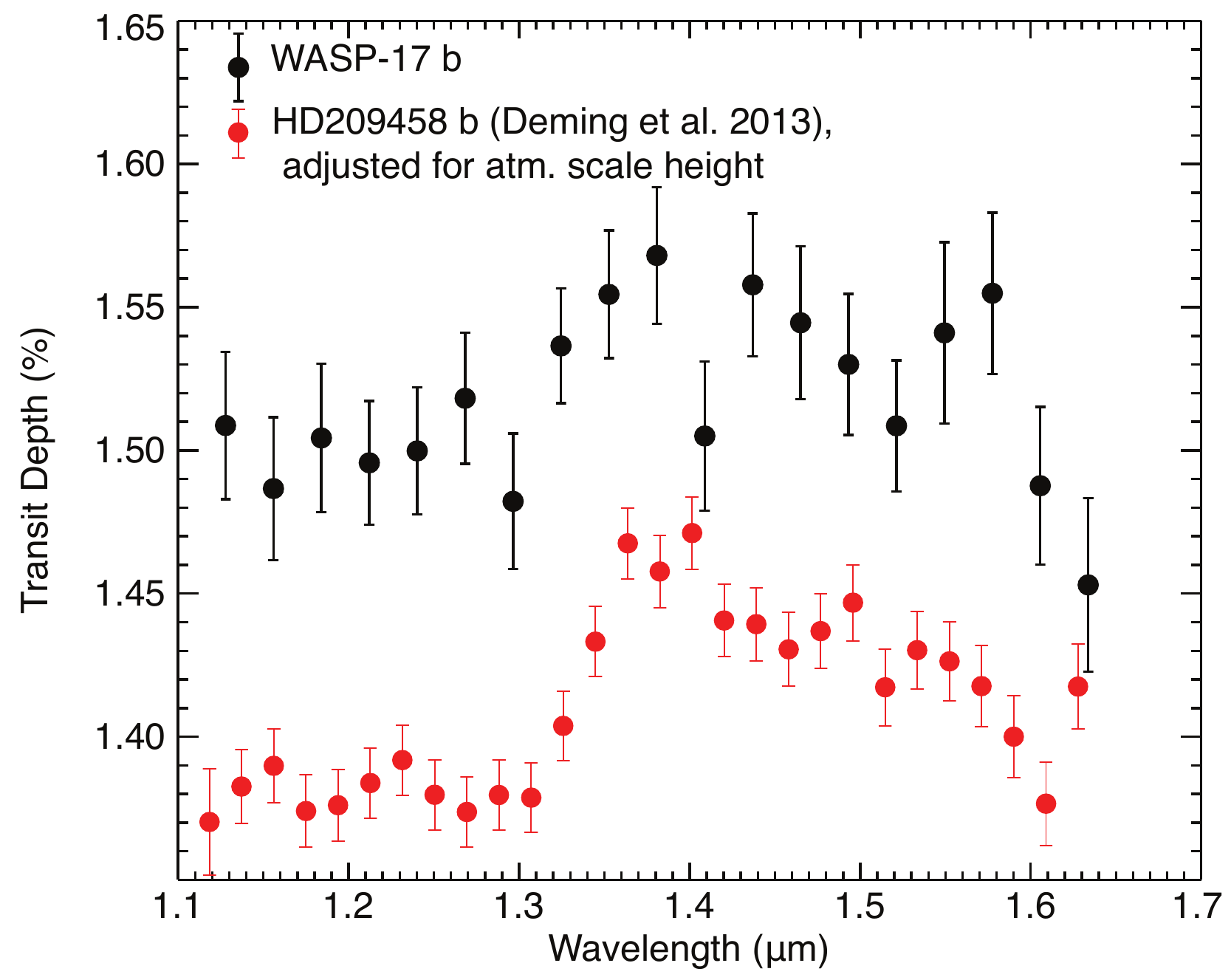}
\caption{Results for WASP-17 b (black) compared with results for HD209458 b from \citet{Deming:2013vu} in red.  The spectrum for HD209458 b has been scaled to compensate for the difference in scale height for the two planets. The spectra match very well, suggesting commonality between the spectra for cooler, smaller planets.}
 \label{fig:deming}
}
\end{figure}

There are no prior spectroscopic analyses of WASP-17 at H-band wavelengths, but we can compare our results with the recent WFC3 observations of HD209458 b by \citet{Deming:2013vu}.  HD209458 b is similar in mass and temperature to WASP-17 b, but with a much smaller scale height - WASP-17 b has a scale height that is 3.4$\times$ larger than HD209458 b.  In Figure~\ref{fig:deming} we plot our spectrum of WASP-17 b with the spectrum of HD209458 b from \citet{Deming:2013vu}, scaled up to compensate for the differences in scale height between the two planets; the spectra match very closely, though there is no evidence for the outlying peak at 1.575\;$\mu$m in the spectrum of HD209458 b.  The similarity between two cooler, lower-mass planets is especially notable considering that dissimilarity between the spectrum for WASP-17 b and the spectra for our other two targets, which are much hotter and more massive.

As stated earlier, we find that the models for WASP-17 b fit best when we include an additional linear slope in the models; we calculate a change in the baseline radius of $\sim$1.63$\times10^{4}$ km across our bandpass for the best-fitting hazy model.  If we assume that this spectral slope is due to a change in effective radius with wavelength due to Rayleigh scattering, we can use Eqn. 4 from \citet{LecavelierDesEtangs:2008he} to compare our spectral slope to similar results for the spectral slope of HD189733 b across optical and IR wavelengths \citep{Pont:2008ft,LecavelierDesEtangs:2008he,Pont:2013ch}.  WASP-17 b is hotter than HD189733 b by $\sim$400K, and the gravity is lower by a factor of $\sim$7; combining these factors leads to a change in altitude across our bandpass of $\sim$4.65$\times10^{3}$ km --10$\times$ larger than for HD189733 b, but still a factor of 3.5$\times$ smaller than our best-fit value.  Considering the lack of a detectable slope in the data for HD209458 b, and the size of the uncertainty bars on our data, we consider this result highly speculative at this point; improved constraints through additional WFC3 observations and/or coincident radius measurements at other wavelengths will be necessary to examine this question in detail.

\subsubsection{WASP-19}
\begin{figure}[htb]
\centering
{
\includegraphics[width=85mm]{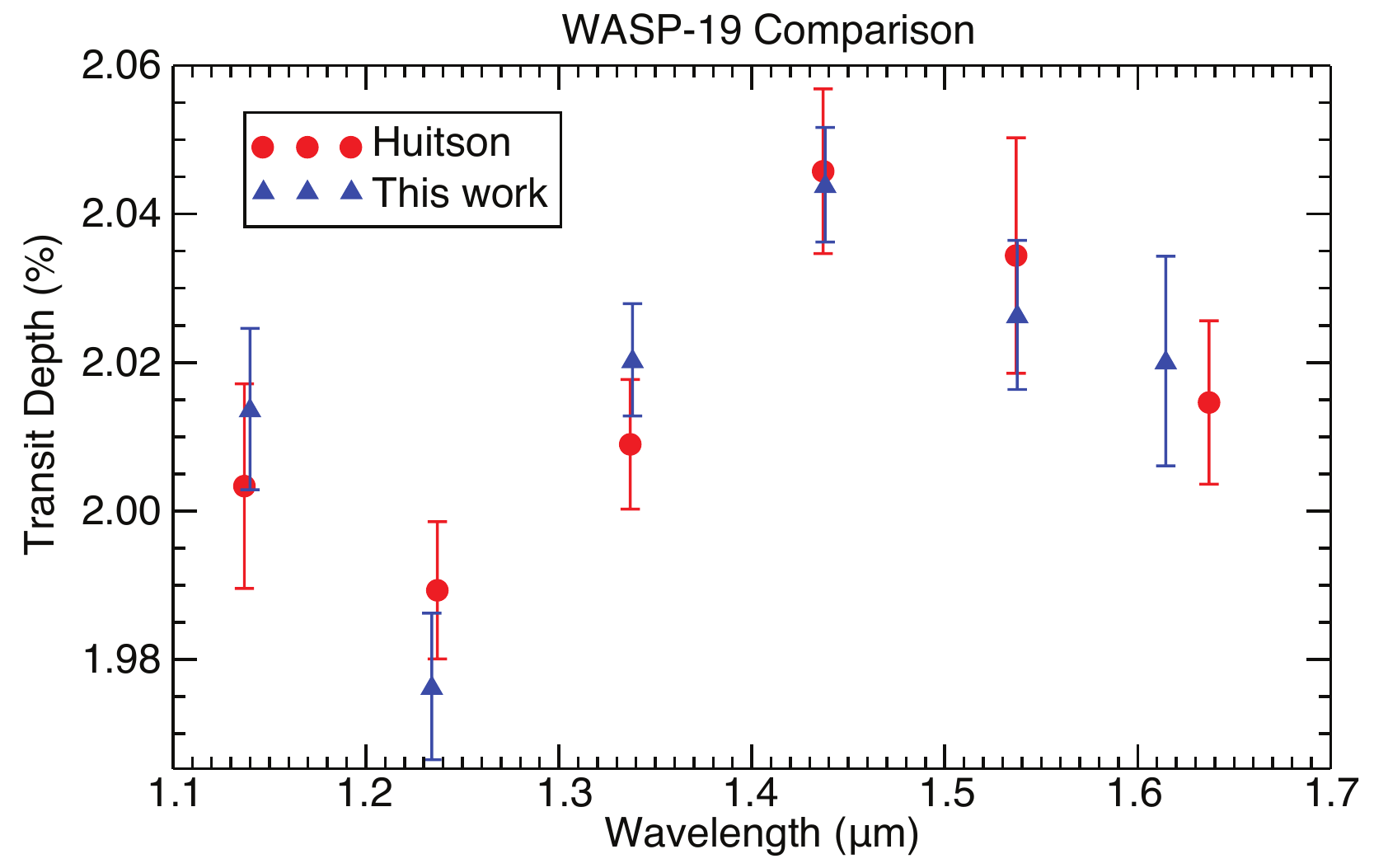}
\caption{Results from \citet{Huitson:2013ud} in red, with results from this work over plotted in blue, binned to the same size, with edges offset due to different choices of spectral binning. The spectra are largely consistent, but comparison with our smaller bin size suggests that the \citet{Huitson:2013ud} may be missing statistically significant features in the spectrum.}
 \label{fig:huitson}
}
\end{figure}

The current data set for WASP-19 was also recently analyzed by \citet{Huitson:2013ud}. Their published results utilized a bin size that is larger than ours by a factor of 3 (0.1\;$\mu$m); they also subtracted the band-integrated residuals from each bin, but then used the divide-oot method on each bin separately and fit for transit depth and a linear trend. In Figure~\ref{fig:huitson} we plot our results using a bin size matched to those of \citet{Huitson:2013ud}. The transit depths using larger bins are well matched to the \citet{Huitson:2013ud} results, but as noted above, with smaller bins we see deviations from the smooth trend that appears to match the lower-resolution results.  \citet{Huitson:2013ud} state that they do not see any major differences beyond increased photon noise when using smaller bin sizes; however, the changes in our spectrum seem to be robust beyond a simple increase in photon noise. \citet{Bean:2013dg} also presented a recent analysis of ground-based transit and occultation observations of WASP-19 at H-band wavelengths. Their results covered the region from 1.25 - 2.4\;$\mu$m, with gaps near the peaks of the water features at 1.37 and 1.9\;$\mu$m.  The analysis of the transit observations yields only four broad bins in our wavelength region, similar in width and position to several of the wavelength bins used by \citet{Huitson:2013ud} and generally consistent with both the Huitson et al. results and our own results for wide bins.

\section{Conclusion}
\label{Con}

In this paper we present our analysis of WFC3 observations of single transits for three exoplanets (WASP-12 b, WASP-17 b and WASP-19 b).  We perform a careful analysis of the band-integrated time series for each target, revealing possible evidence of curvature in the out-of-transit data for WASP-12 and WASP-19 and evidence for a star spot in the light curve for WASP-19.  We confirm that the repeating ramp-like or hook-like artifacts seen in a number of observations of exoplanets with WFC3 (which we call the ``buffer-ramp") can be removed in the band-integrated light curve using the divide-oot method from \citet{Berta:2012ff}, but we develop an alternate method for removing the various systematic trends in the individual channels or bins of multiple channels that utilizes the residuals of the fit to the band-integrated light curve as well as measurements of the vertical and horizontal shift of the spectrum on the detector over time.  We utilize a model selection strategy that relies on the Bayesian Information Criterion to determine the significance of fitting for individual systematic components, allowing us to identify trends due to changes in the amplitude of the buffer-ramp and the impact of spectral shifts on the flux in individual spectral bins.  We present final transit spectra for each exoplanet using 0.027\;$\mu$m channel bins, and argue that this is the optimal bin size for increasing S/N while avoiding any loss of spectral information that exceeds the photon-noise limit.  When we use similar binning sizes to those used in previous analyses of the data for WASP-12 \citep{Swain:2013hn,Stevenson:2013wf} and WASP-19 \citep{Huitson:2013ud}, we can reproduce the earlier results to within uncertainties except for the shortest-wavelength bin for WASP-12; this discrepancy may be due to treatment of data that falls on the steep spectral slope of the WFC3 sensitivity curve.

Our analysis demonstrates that precisions close to the photon-noise limit are possible for measurements of wavelength-dependent transit depths with WFC3 with the observation of only a single transit event even for relatively dim targets ($H>10.2$). Measurements of the absolute transit depth are fundamentally limited by our ability to constrain parameters such as limb darkening and mid-transit time, and the phasing of HST orbits across the light curve has a significant impact on our final uncertainties in $R_{p}/R_{*}$ for our band-integrated light curves. However, using our transit model including systematic trends, we show that the uncertainties for individual bins are not strongly affected by the light curve sampling and depend only on the number of photons acquired in transit and out-of-transit.  Future observations of these targets that utilize the newly implemented spatial scan mode will allow for increased efficiency and improved sensitivity.

Comparison with theoretical models by \citet{Burrows:2008ff} and \citet{Madhusudhan:2012ga} strongly suggest the presence of water absorption between 1.4\,$\mu$m and 1.55\,$\mu$m in WASP-17 b, and models with the inclusion of haze fit the data better than models without haze.  For WASP-12 b and WASP-19 b the agreement with standard models including water absorption is not as clear. In particular, the spectral region beyond 1.45\;$\mu$m shows increased absorption for all our targets beyond what is predicted from water-rich models; carbon-rich models provide a better match in this region, but significant discrepancies remain.  We therefore believe that firm conclusions on atmospheric composition are impossible without more sensitive observations and/or a full analysis of multi-wavelength data at both optical and NIR wavelengths.

\acknowledgements


\bibliographystyle{apj}

\begin{thebibliography}{71}
\expandafter\ifx\csname natexlab\endcsname\relax\def\natexlab#1{#1}\fi

\bibitem[{Albrecht {et~al.}(2012)Albrecht, Winn, Johnson, Howard, Marcy,
  Butler, Arriagada, Crane, Shectman, Thompson, Hirano, Bakos, \&
  Hartman}]{Albrecht:2012hk}
Albrecht, S., Winn, J.~N., Johnson, J.~A., Howard, A.~W., Marcy, G.~W., Butler,
  R.~P., Arriagada, P., Crane, J.~D., Shectman, S.~A., Thompson, I.~B., Hirano,
  T., Bakos, G., \& Hartman, J.~D. 2012, The Astrophysical Journal, 757, 18

\bibitem[{Anderson {et~al.}(2010)Anderson, Hellier, Gillon, Triaud, Smalley,
  Hebb, Cameron, Maxted, Queloz, West, Bentley, Enoch, Horne, Lister, Mayor,
  Parley, Pepe, Pollacco, S{\'e}gransan, Udry, \& Wilson}]{Anderson:2010dx}
Anderson, D.~R., Hellier, C., Gillon, M., Triaud, A. H. M.~J., Smalley, B.,
  Hebb, L., Cameron, A.~C., Maxted, P. F.~L., Queloz, D., West, R.~G., Bentley,
  S.~J., Enoch, B., Horne, K., Lister, T.~A., Mayor, M., Parley, N.~R., Pepe,
  F., Pollacco, D., S{\'e}gransan, D., Udry, S., \& Wilson, D.~M. 2010, The
  Astrophysical Journal, 709, 159

\bibitem[{Anderson {et~al.}(2011)Anderson, Smith, Lanotte, Barman, Cameron,
  Campo, Gillon, Harrington, Hellier, Maxted, Queloz, Triaud, \&
  Wheatley}]{Anderson:2011hf}
Anderson, D.~R., Smith, A. M.~S., Lanotte, A.~A., Barman, T.~S., Cameron,
  A.~C., Campo, C.~J., Gillon, M., Harrington, J., Hellier, C., Maxted, P.
  F.~L., Queloz, D., Triaud, A. H. M.~J., \& Wheatley, P.~J. 2011, Monthly
  Notices of the Royal Astronomical Society, 416, 2108

\bibitem[{Anderson {et~al.}(2013)Anderson, Smith, Madhusudhan, Wheatley,
  Collier-Cameron, Hellier, Campo, Gillon, Harrington, Maxted, Pollacco,
  Queloz, Smalley, Triaud, \& West}]{Anderson:2013dx}
Anderson, D.~R., Smith, A. M.~S., Madhusudhan, N., Wheatley, P.~J.,
  Collier-Cameron, A., Hellier, C., Campo, C., Gillon, M., Harrington, J.,
  Maxted, P. F.~L., Pollacco, D., Queloz, D., Smalley, B., Triaud, A. H. M.~J.,
  \& West, R.~G. 2013, Monthly Notices of the Royal Astronomical Society, 430,
  3422

\bibitem[{Batygin \& Stevenson(2010)}]{Batygin:2010dz}
Batygin, K. \& Stevenson, D.~J. 2010, The Astrophysical Journal, 714, L238

\bibitem[{Bayliss {et~al.}(2010)Bayliss, Winn, Mardling, \&
  Sackett}]{Bayliss:2010ch}
Bayliss, D. D.~R., Winn, J.~N., Mardling, R.~A., \& Sackett, P.~D. 2010, The
  Astrophysical Journal, 722, L224

\bibitem[{Bean {et~al.}(2013)Bean, Desert, Seifahrt, Madhusudhan, Chilingarian,
  Homeier, \& Szentgyorgyi}]{Bean:2013dg}
Bean, J.~L., Desert, J.-M., Seifahrt, A., Madhusudhan, N., Chilingarian, I.,
  Homeier, D., \& Szentgyorgyi, A. 2013, The Astrophysical Journal, 771, 108

\bibitem[{Bechter {et~al.}(2013)Bechter, Crepp, Ngo, Knutson, Batygin, Hinkley,
  Muirhead, Johnson, Howard, Montet, Matthews, \& Morton}]{Bechter:2013uv}
Bechter, E.~B., Crepp, J.~R., Ngo, H., Knutson, H.~A., Batygin, K., Hinkley,
  S., Muirhead, P.~S., Johnson, J.~A., Howard, A.~W., Montet, B.~T., Matthews,
  C.~T., \& Morton, T.~D. 2013, arXiv.org, 6857

\bibitem[{Bergfors {et~al.}(2013)Bergfors, Brandner, Daemgen, Biller, Hippler,
  Janson, Kudryavtseva, Gei{\ss}ler, Henning, \& K{\"o}hler}]{Bergfors:2013de}
Bergfors, C., Brandner, W., Daemgen, S., Biller, B., Hippler, S., Janson, M.,
  Kudryavtseva, N., Gei{\ss}ler, K., Henning, T., \& K{\"o}hler, R. 2013,
  Monthly Notices of the Royal Astronomical Society, 428, 182

\bibitem[{Berta {et~al.}(2012)Berta, Charbonneau, Desert, Miller-Ricci~Kempton,
  McCullough, Burke, Fortney, Irwin, Nutzman, \& Homeier}]{Berta:2012ff}
Berta, Z.~K., Charbonneau, D., Desert, J.-M., Miller-Ricci~Kempton, E.,
  McCullough, P.~R., Burke, C.~J., Fortney, J.~J., Irwin, J., Nutzman, P., \&
  Homeier, D. 2012, The Astrophysical Journal, 747, 35

\bibitem[{Bertin \& Arnouts(1996)}]{Bertin:1996ww}
Bertin, E. \& Arnouts, S. 1996, Astronomy and Astrophysics Supplement, 117, 393

\bibitem[{Bodenheimer {et~al.}(2001)Bodenheimer, Lin, \&
  Mardling}]{Bodenheimer:2001eu}
Bodenheimer, P., Lin, D. N.~C., \& Mardling, R.~A. 2001, The Astrophysical
  Journal, 548, 466

\bibitem[{Burrows {et~al.}(2008)Burrows, Budaj, \& Hubeny}]{Burrows:2008ff}
Burrows, A., Budaj, J., \& Hubeny, I. 2008, The Astrophysical Journal, 678,
  1436

\bibitem[{Burrows {et~al.}(2000)Burrows, Guillot, Hubbard, Marley, Saumon,
  Lunine, \& Sudarsky}]{Burrows:2000ho}
Burrows, A., Guillot, T., Hubbard, W.~B., Marley, M.~S., Saumon, D., Lunine,
  J.~I., \& Sudarsky, D. 2000, The Astrophysical Journal, 534, L97

\bibitem[{Burrows {et~al.}(2006)Burrows, Sudarsky, \& Hubeny}]{Burrows:2006ga}
Burrows, A., Sudarsky, D., \& Hubeny, I. 2006, The Astrophysical Journal, 650,
  1140

\bibitem[{Campo {et~al.}(2011)Campo, Harrington, Hardy, Stevenson, Nymeyer,
  Ragozzine, Lust, Anderson, Collier-Cameron, Blecic, Britt, Bowman, Wheatley,
  Loredo, Deming, Hebb, Hellier, Maxted, Pollaco, \& West}]{Campo:2011fx}
Campo, C.~J., Harrington, J., Hardy, R.~A., Stevenson, K.~B., Nymeyer, S.,
  Ragozzine, D., Lust, N.~B., Anderson, D.~R., Collier-Cameron, A., Blecic, J.,
  Britt, C. B.~T., Bowman, W.~C., Wheatley, P.~J., Loredo, T.~J., Deming, D.,
  Hebb, L., Hellier, C., Maxted, P. F.~L., Pollaco, D., \& West, R.~G. 2011,
  The Astrophysical Journal, 727, 125

\bibitem[{Castelli \& Kurucz(2004)}]{Castelli:2004ti}
Castelli, F. \& Kurucz, R.~L. 2004, arXiv.org, 5087

\bibitem[{Charbonneau {et~al.}(2005)Charbonneau, Allen, Megeath, Torres,
  Alonso, Brown, Gilliland, Latham, Mandushev, O'Donovan, \&
  Sozzetti}]{Charbonneau:2005be}
Charbonneau, D., Allen, L.~E., Megeath, S.~T., Torres, G., Alonso, R., Brown,
  T.~M., Gilliland, R.~L., Latham, D.~W., Mandushev, G., O'Donovan, F.~T., \&
  Sozzetti, A. 2005, The Astrophysical Journal, 626, 523

\bibitem[{Charbonneau {et~al.}(2002)Charbonneau, Brown, Noyes, \&
  Gilliland}]{Charbonneau:2002er}
Charbonneau, D., Brown, T.~M., Noyes, R.~W., \& Gilliland, R.~L. 2002, The
  Astrophysical Journal, 568, 377

\bibitem[{Claret \& Bloemen(2011)}]{Claret:2011gy}
Claret, A. \& Bloemen, S. 2011, Astronomy and Astrophysics, 529, 75

\bibitem[{Cowan \& Agol(2011)}]{Cowan:2011kw}
Cowan, N.~B. \& Agol, E. 2011, The Astrophysical Journal, 726, 82

\bibitem[{Cowan {et~al.}(2012)Cowan, Machalek, Croll, Shekhtman, Burrows,
  Deming, Greene, \& Hora}]{Cowan:2012gp}
Cowan, N.~B., Machalek, P., Croll, B., Shekhtman, L.~M., Burrows, A., Deming,
  D., Greene, T., \& Hora, J.~L. 2012, The Astrophysical Journal, 747, 82

\bibitem[{Croll {et~al.}(2011)Croll, Lafreniere, Albert, Jayawardhana, Fortney,
  \& Murray}]{Croll:2011jt}
Croll, B., Lafreniere, D., Albert, L., Jayawardhana, R., Fortney, J.~J., \&
  Murray, N. 2011, The Astronomical Journal, 141, 30

\bibitem[{Crossfield {et~al.}(2012)Crossfield, Barman, Hansen, Tanaka, \&
  Kodama}]{Crossfield:2012gm}
Crossfield, I. J.~M., Barman, T., Hansen, B. M.~S., Tanaka, I., \& Kodama, T.
  2012, The Astrophysical Journal, 760, 140

\bibitem[{Deming {et~al.}(2005)Deming, Seager, Richardson, \&
  Harrington}]{Deming:2005fg}
Deming, D., Seager, S., Richardson, L.~J., \& Harrington, J. 2005, Nature, 434,
  740

\bibitem[{Deming {et~al.}(2013)Deming, Wilkins, McCullough, Burrows, Fortney,
  Agol, Dobbs-Dixon, Madhusudhan, Crouzet, Desert, Gilliland, Haynes, Knutson,
  Line, Magic, Mandell, Ranjan, Charbonneau, Clampin, Seager, \&
  Showman}]{Deming:2013vu}
Deming, D., Wilkins, A., McCullough, P., Burrows, A., Fortney, J.~J., Agol, E.,
  Dobbs-Dixon, I., Madhusudhan, N., Crouzet, N., Desert, J.-M., Gilliland,
  R.~L., Haynes, K., Knutson, H.~A., Line, M., Magic, Z., Mandell, A.~M.,
  Ranjan, S., Charbonneau, D., Clampin, M., Seager, S., \& Showman, A.~P. 2013,
  arXiv.org, 1141

\bibitem[{Deming {et~al.}(2012)Deming, Wilkins, McCullough, Madhusudhan, Agol,
  Burrows, Charbonneau, Clampin, Desert, Gilliland, Knutson, Mandell, Ranjan,
  Seager, \& Showman}]{Deming:2012vd}
Deming, D., Wilkins, A., McCullough, P., Madhusudhan, N., Agol, E., Burrows,
  A., Charbonneau, D., Clampin, M., Desert, J., Gilliland, R., Knutson, H.,
  Mandell, A., Ranjan, S., Seager, S., \& Showman, A. 2012, American
  Astronomical Society, 219

\bibitem[{Dressel(2012)}]{Dressel2012handbook}
Dressel, L. 2012, {Wide Field Camera 3 Instrument Handbook, Version 5.0}
  (Baltimore: STScI)

\bibitem[{Fabrycky \& Tremaine(2007)}]{Fabrycky:2007jh}
Fabrycky, D. \& Tremaine, S. 2007, The Astrophysical Journal, 669, 1298

\bibitem[{Ford(2005)}]{Ford:2005gi}
Ford, E.~B. 2005, The Astronomical Journal, 129, 1706

\bibitem[{Fortney {et~al.}(2008)Fortney, Lodders, Marley, \&
  Freedman}]{Fortney:2008ce}
Fortney, J.~J., Lodders, K., Marley, M.~S., \& Freedman, R.~S. 2008, The
  Astrophysical Journal, 678, 1419

\bibitem[{Grillmair {et~al.}(2008)Grillmair, Burrows, Charbonneau, Armus,
  Stauffer, Meadows, Cleve, Braun, \& Levine}]{Grillmair:2008je}
Grillmair, C.~J., Burrows, A., Charbonneau, D., Armus, L., Stauffer, J.,
  Meadows, V., Cleve, J.~v., Braun, K.~v., \& Levine, D. 2008, Nature, 456, 767

\bibitem[{Grillmair {et~al.}(2007)Grillmair, Charbonneau, Burrows, Armus,
  Stauffer, Meadows, Van~Cleve, \& Levine}]{Grillmair:2007ee}
Grillmair, C.~J., Charbonneau, D., Burrows, A., Armus, L., Stauffer, J.,
  Meadows, V., Van~Cleve, J., \& Levine, D. 2007, The Astrophysical Journal,
  658, L115

\bibitem[{Guillot \& Showman(2002)}]{Guillot:2002cq}
Guillot, T. \& Showman, A.~P. 2002, Astronomy and Astrophysics, 385, 156

\bibitem[{Hebb {et~al.}(2009)Hebb, Collier-Cameron, Loeillet, Pollacco,
  H{\'e}brard, Street, Bouchy, Stempels, Moutou, Simpson, Udry, Joshi, West,
  Skillen, Wilson, McDonald, Gibson, Aigrain, Anderson, Benn, Christian, Enoch,
  Haswell, Hellier, Horne, Irwin, Lister, Maxted, Mayor, Norton, Parley, Pont,
  Queloz, Smalley, \& Wheatley}]{Hebb:2009fw}
Hebb, L., Collier-Cameron, A., Loeillet, B., Pollacco, D., H{\'e}brard, G.,
  Street, R.~A., Bouchy, F., Stempels, H.~C., Moutou, C., Simpson, E., Udry,
  S., Joshi, Y.~C., West, R.~G., Skillen, I., Wilson, D.~M., McDonald, I.,
  Gibson, N.~P., Aigrain, S., Anderson, D.~R., Benn, C.~R., Christian, D.~J.,
  Enoch, B., Haswell, C.~A., Hellier, C., Horne, K., Irwin, J., Lister, T.~A.,
  Maxted, P., Mayor, M., Norton, A.~J., Parley, N., Pont, F., Queloz, D.,
  Smalley, B., \& Wheatley, P.~J. 2009, The Astrophysical Journal, 693, 1920

\bibitem[{Hellier {et~al.}(2011)Hellier, Anderson, Collier-Cameron, Miller,
  Queloz, Smalley, Southworth, \& Triaud}]{Hellier:2011hf}
Hellier, C., Anderson, D.~R., Collier-Cameron, A., Miller, G. R.~M., Queloz,
  D., Smalley, B., Southworth, J., \& Triaud, A. H. M.~J. 2011, The
  Astrophysical Journal, 730, L31

\bibitem[{Howe \& Burrows(2012)}]{Howe:2012et}
Howe, A.~R. \& Burrows, A.~S. 2012, The Astrophysical Journal, 756, 176

\bibitem[{Hubeny {et~al.}(2003)Hubeny, Burrows, \& Sudarsky}]{Hubeny:2003eb}
Hubeny, I., Burrows, A., \& Sudarsky, D. 2003, The Astrophysical Journal, 594,
  1011

\bibitem[{Huitson {et~al.}(2013)Huitson, Sing, Pont, Fortney, Burrows, Wilson,
  Ballester, Nikolov, Gibson, Deming, Aigrain, Evans, Henry, Lecavelier
  Des~Etangs, Showman, Vidal-Madjar, \& Zahnle}]{Huitson:2013ud}
Huitson, C.~M., Sing, D.~K., Pont, F., Fortney, J.~J., Burrows, A.~S., Wilson,
  P.~A., Ballester, G.~E., Nikolov, N., Gibson, N.~P., Deming, D., Aigrain, S.,
  Evans, T.~M., Henry, G.~W., Lecavelier Des~Etangs, A., Showman, A.~P.,
  Vidal-Madjar, A., \& Zahnle, K. 2013, arXiv.org, 2083

\bibitem[{Ibgui \& Burrows(2009)}]{Ibgui:2009fw}
Ibgui, L. \& Burrows, A. 2009, The Astrophysical Journal, 700, 1921

\bibitem[{Jackson {et~al.}(2008)Jackson, Greenberg, \& Barnes}]{Jackson:2008ip}
Jackson, B., Greenberg, R., \& Barnes, R. 2008, The Astrophysical Journal, 678,
  1396

\bibitem[{Knutson {et~al.}(2007)Knutson, Charbonneau, Allen, Fortney, Agol,
  Cowan, Showman, Cooper, \& Megeath}]{Knutson:2007bl}
Knutson, H.~A., Charbonneau, D., Allen, L.~E., Fortney, J.~J., Agol, E., Cowan,
  N.~B., Showman, A.~P., Cooper, C.~S., \& Megeath, S.~T. 2007, Nature, 447,
  183

\bibitem[{K{\"u}mmel {et~al.}(2009)K{\"u}mmel, Walsh, Pirzkal, Kuntschner, \&
  Pasquali}]{Kummel:2009dn}
K{\"u}mmel, M., Walsh, J.~R., Pirzkal, N., Kuntschner, H., \& Pasquali, A.
  2009, Publications of the Astronomical Society of the Pacific, 121, 59

\bibitem[{Kuntschner {et~al.}(2008)Kuntschner, Bushouse, Walsh, \&
  K{\"u}mmel}]{Kuntschner:2008tq}
Kuntschner, H., Bushouse, H., Walsh, J.~R., \& K{\"u}mmel, M. 2008, ST-ECF
  Instrument Science Report WFC3-2008-16, 16

\bibitem[{Lecavelier Des~Etangs {et~al.}(2008)Lecavelier Des~Etangs, Pont,
  Vidal-Madjar, \& Sing}]{LecavelierDesEtangs:2008he}
Lecavelier Des~Etangs, A., Pont, F., Vidal-Madjar, A., \& Sing, D. 2008,
  Astronomy and Astrophysics, 481, L83

\bibitem[{Lendl {et~al.}(2013)Lendl, Gillon, Queloz, Alonso, Fumel, Jehin, \&
  Naef}]{Lendl:2013jda}
Lendl, M., Gillon, M., Queloz, D., Alonso, R., Fumel, A., Jehin, E., \& Naef,
  D. 2013, Astronomy and Astrophysics, 552, 2

\bibitem[{Liddle(2004)}]{Liddle:2004cz}
Liddle, A.~R. 2004, Monthly Notices of the Royal Astronomical Society, 351, L49

\bibitem[{Maciejewski {et~al.}(2013)Maciejewski, Dimitrov, Seeliger, Raetz,
  Bukowiecki, Kitze, Errmann, Nowak, Niedzielski, Popov, Marka,
  Go{\'z}dziewski, Neuh{\"a}user, Ohlert, Hinse, Lee, Lee, Yoon, Berndt,
  Gilbert, Ginski, Hohle, Mugrauer, R{\"o}ll, Schmidt, Tetzlaff, Mancini,
  Southworth, Dall'Ora, Ciceri, Zambelli, Corfini, Takahashi, Tachihara,
  Benk{\H o}, S{\'a}rneczky, Szabo, Varga, Va{\v n}ko, Joshi, \&
  Chen}]{Maciejewski:2013ip}
Maciejewski, G., Dimitrov, D., Seeliger, M., Raetz, S., Bukowiecki, Å., Kitze,
  M., Errmann, R., Nowak, G., Niedzielski, A., Popov, V., Marka, C.,
  Go{\'z}dziewski, K., Neuh{\"a}user, R., Ohlert, J., Hinse, T.~C., Lee, J.~W.,
  Lee, C.~U., Yoon, J.~N., Berndt, A., Gilbert, H., Ginski, C., Hohle, M.~M.,
  Mugrauer, M., R{\"o}ll, T., Schmidt, T. O.~B., Tetzlaff, N., Mancini, L.,
  Southworth, J., Dall'Ora, M., Ciceri, S., Zambelli, R., Corfini, G.,
  Takahashi, H., Tachihara, K., Benk{\H o}, J.~M., S{\'a}rneczky, K., Szabo,
  G.~M., Varga, T.~N., Va{\v n}ko, M., Joshi, Y.~C., \& Chen, W.~P. 2013,
  Astronomy and Astrophysics, 551, 108

\bibitem[{Madhusudhan(2012)}]{Madhusudhan:2012ga}
Madhusudhan, N. 2012, The Astrophysical Journal, 758, 36

\bibitem[{Madhusudhan {et~al.}(2011)Madhusudhan, Harrington, Stevenson,
  Nymeyer, Campo, Wheatley, Deming, Blecic, Hardy, Lust, Anderson,
  Collier-Cameron, Britt, Bowman, Hebb, Hellier, Maxted, Pollacco, \&
  West}]{Madhusudhan:2011kw}
Madhusudhan, N., Harrington, J., Stevenson, K.~B., Nymeyer, S., Campo, C.~J.,
  Wheatley, P.~J., Deming, D., Blecic, J., Hardy, R.~A., Lust, N.~B., Anderson,
  D.~R., Collier-Cameron, A., Britt, C. B.~T., Bowman, W.~C., Hebb, L.,
  Hellier, C., Maxted, P. F.~L., Pollacco, D., \& West, R.~G. 2011, Nature,
  469, 64

\bibitem[{Madhusudhan \& Seager(2009)}]{Madhusudhan:2009gd}
Madhusudhan, N. \& Seager, S. 2009, The Astrophysical Journal, 707, 24

\bibitem[{Mandel \& Agol(2002)}]{Mandel:2002bb}
Mandel, K. \& Agol, E. 2002, The Astrophysical Journal, 580, L171

\bibitem[{McCullough \& Deustua(2008)}]{McCullough:2008tl}
McCullough, P. \& Deustua, S. 2008, Instrument Science Report WFC3 2008-33, 33

\bibitem[{McCullough \& MacKenty(2012)}]{2012McCullough_spatscan}
McCullough, P. \& MacKenty, J. 2012, Instrument Science Report WFC3 2012-08, 8

\bibitem[{Pont {et~al.}(2008)Pont, Knutson, Gilliland, Moutou, \&
  Charbonneau}]{Pont:2008ft}
Pont, F., Knutson, H., Gilliland, R.~L., Moutou, C., \& Charbonneau, D. 2008,
  Monthly Notices of the Royal Astronomical Society, 385, 109

\bibitem[{Pont {et~al.}(2013)Pont, Sing, Gibson, Aigrain, Henry, \&
  Husnoo}]{Pont:2013ch}
Pont, F., Sing, D.~K., Gibson, N.~P., Aigrain, S., Henry, G., \& Husnoo, N.
  2013, Monthly Notices of the Royal Astronomical Society, 432, 2917

\bibitem[{Rasio \& Ford(1996)}]{Rasio:1996ie}
Rasio, F.~A. \& Ford, E.~B. 1996, Science, 274, 954

\bibitem[{Richardson {et~al.}(2007)Richardson, Deming, Horning, Seager, \&
  Harrington}]{Richardson:2007eu}
Richardson, L.~J., Deming, D., Horning, K., Seager, S., \& Harrington, J. 2007,
  Nature, 445, 892

\bibitem[{Rogers {et~al.}(2012)Rogers, Lin, \& Lau}]{Rogers:2012dm}
Rogers, T.~M., Lin, D. N.~C., \& Lau, H. H.~B. 2012, The Astrophysical Journal,
  1209, 2435

\bibitem[{Schwarz(1978)}]{Schwarz:1978kf}
Schwarz, G. 1978, The Annals of Statistics, 6, 461

\bibitem[{Smith {et~al.}(2008)Smith, Zavodny, Rahmer, \& Bonati}]{Smith:2008by}
Smith, R.~M., Zavodny, M., Rahmer, G., \& Bonati, M. 2008, High Energy, 7021,
  15

\bibitem[{Southworth {et~al.}(2012)Southworth, Hinse, Dominik, Fang, Harpsoe,
  Jorgensen, Kerins, Liebig, Mancini, Skottfelt, Anderson, Smalley,
  Tregloan-Reed, Wertz, Alsubai, Bozza, Calchi~Novati, Dreizler, Gu,
  Hundertmark, Jessen-Hansen, Kains, Kjeldsen, Lund, Lundkvist, Mathiasen,
  Penny, Rahvar, Ricci, Scarpetta, Snodgrass, \& Surdej}]{Southworth:2012fv}
Southworth, J., Hinse, T.~C., Dominik, M., Fang, X.~S., Harpsoe, K., Jorgensen,
  U.~G., Kerins, E., Liebig, C., Mancini, L., Skottfelt, J., Anderson, D.~R.,
  Smalley, B., Tregloan-Reed, J., Wertz, O., Alsubai, K.~A., Bozza, V.,
  Calchi~Novati, S., Dreizler, S., Gu, S.~H., Hundertmark, M., Jessen-Hansen,
  J., Kains, N., Kjeldsen, H., Lund, M.~N., Lundkvist, M., Mathiasen, M.,
  Penny, M.~T., Rahvar, S., Ricci, D., Scarpetta, G., Snodgrass, C., \& Surdej,
  J. 2012, Monthly Notices of the Royal Astronomical Society, 426, 1338

\bibitem[{Spiegel {et~al.}(2009)Spiegel, Silverio, \& Burrows}]{Spiegel:2009ja}
Spiegel, D.~S., Silverio, K., \& Burrows, A. 2009, The Astrophysical Journal,
  699, 1487

\bibitem[{Stevenson {et~al.}(2013)Stevenson, Bean, Seifahrt, Desert,
  Madhusudhan, Bergmann, Kreidberg, \& Homeier}]{Stevenson:2013wf}
Stevenson, K.~B., Bean, J.~L., Seifahrt, A., Desert, J.-M., Madhusudhan, N.,
  Bergmann, M., Kreidberg, L., \& Homeier, D. 2013, arXiv.org, 1670

\bibitem[{Swain {et~al.}(2013)Swain, Deroo, Tinetti, Hollis, Tessenyi, Line,
  Kawahara, Fujii, Showman, \& Yurchenko}]{Swain:2013hn}
Swain, M., Deroo, P., Tinetti, G., Hollis, M., Tessenyi, M., Line, M.,
  Kawahara, H., Fujii, Y., Showman, A.~P., \& Yurchenko, S.~N. 2013, Icarus,
  225, 432

\bibitem[{Swain {et~al.}(2009{\natexlab{a}})Swain, Tinetti, Vasisht, Deroo,
  Griffith, Bouwman, Chen, Yung, Burrows, Brown, Matthews, Rowe, Kuschnig, \&
  Angerhausen}]{Swain:2009ed}
Swain, M.~R., Tinetti, G., Vasisht, G., Deroo, P., Griffith, C., Bouwman, J.,
  Chen, P., Yung, Y., Burrows, A., Brown, L.~R., Matthews, J., Rowe, J.~F.,
  Kuschnig, R., \& Angerhausen, D. 2009{\natexlab{a}}, The Astrophysical
  Journal, 704, 1616

\bibitem[{Swain {et~al.}(2008)Swain, Vasisht, \& Tinetti}]{Swain:2008fm}
Swain, M.~R., Vasisht, G., \& Tinetti, G. 2008, Nature, 452, 329

\bibitem[{Swain {et~al.}(2009{\natexlab{b}})Swain, Vasisht, Tinetti, Bouwman,
  Chen, Yung, Deming, \& Deroo}]{Swain:2009ds}
Swain, M.~R., Vasisht, G., Tinetti, G., Bouwman, J., Chen, P., Yung, Y.,
  Deming, D., \& Deroo, P. 2009{\natexlab{b}}, The Astrophysical Journal, 690,
  L114

\bibitem[{Tregloan-Reed {et~al.}(2013)Tregloan-Reed, Southworth, \&
  Tappert}]{TregloanReed:2013gd}
Tregloan-Reed, J., Southworth, J., \& Tappert, C. 2013, Monthly Notices of the
  Royal Astronomical Society, 428, 3671

\bibitem[{Triaud {et~al.}(2010)Triaud, Collier-Cameron, Queloz, Anderson,
  Gillon, Hebb, Hellier, Loeillet, Maxted, Mayor, Pepe, Pollacco, Segransan,
  Smalley, Udry, West, \& Wheatley}]{Triaud:2010hr}
Triaud, A. H. M.~J., Collier-Cameron, A., Queloz, D., Anderson, D.~R., Gillon,
  M., Hebb, L., Hellier, C., Loeillet, B., Maxted, P. F.~L., Mayor, M., Pepe,
  F., Pollacco, D., Segransan, D., Smalley, B., Udry, S., West, R.~G., \&
  Wheatley, P.~J. 2010, Astronomy and Astrophysics, 524, 25

\bibitem[{Weidenschilling \& Marzari(1996)}]{Weidenschilling:1996ig}
Weidenschilling, S.~J. \& Marzari, F. 1996, Nature, 384, 619

\end{thebibliography}

\end{document}